\documentclass[twocolumn]{aastex701}
\usepackage{CJKutf8} 

\submitjournal{ApJ}

\shortauthors{Zaritsky, et al.}
\shorttitle{A Gyr in the Life}

\begin{document}

\title{Systematically Measuring Ultra-Diffuse Galaxies. IX. A Gyr in the Life of Nearby Low Surface Brightness Galaxies}
  
\correspondingauthor{Dennis Zaritsky}
\email{dennis.zaritsky@gmail.com}

\author[0000-0002-5177-727X]{Dennis Zaritsky}
\affiliation{Steward Observatory and Department of Astronomy, University of Arizona, 933 N. Cherry Ave., Tucson, AZ 85721, USA}
\email{dennis.zaritsky@gmail.com}

\author[0000-0001-7618-8212]{Richard Donnerstein}
\affiliation{Steward Observatory and Department of Astronomy, University of Arizona, 933 N. Cherry Ave., Tucson, AZ 85721, USA}
\email{rdonnest@arizona.edu}

\author[0000-0001-8568-8729]{Loraine Sandoval Ascencio}
\affiliation{Department of Physics \& Astronomy, University of California, Irvine, 4129 Reines Hall, Irvine, CA 92697, USA} 
\email{lorainas@uci.edu}

\author[0000-0003-1371-6019]{M.C. Cooper}
\affiliation{Department of Physics \& Astronomy, University of California, Irvine, 4129 Reines Hall, Irvine, CA 92697, USA} 
\email{cooper@uci.edu}

\author[0000-0002-7013-4392]{Donghyeon J. Khim}
\affiliation{Steward Observatory and Department of Astronomy, University of Arizona, 933 N. Cherry Ave., Tucson, AZ 85721, USA}
\email{galaxydiver@arizona.edu}

\author[0000-0002-0956-7949]{Kristine Spekkens}
\affiliation{Department of Physics, Engineering Physics and Astronomy Queen's University Kingston, ON K7L 3N6, Canada}
\email{kristine.spekkens@queensu.ca}

\begin{abstract}
We augment the published optical photometry of ultra-diffuse galaxy candidates in the SMUDGes catalog with  UV and IR  measurements to investigate the recent ($<1$ Gyr) star formation history of 966 galaxies. We find that 1) we classify star forming, post-starforming, and quenched galaxies with a precision that is comparable  to that of spectroscopic studies, 2) the star forming systems are sub-normally efficient and would not have formed their current stellar mass at their current star formation rate over a Hubble time, 3) the sample is biased against more strongly star forming systems by the central surface brightness criterion of ultra-diffuse galaxies, 4) for  galaxies that are not quenched, the timescale of star formation episodes in this sample is typically $\lesssim$ 1 Gyr, 5) post-starburst galaxies in the sample tend to be of lower stellar mass and star forming galaxies of higher stellar mass, suggesting that the star forming behavior of these galaxies does depend on mass, and 6) there is a marginal indication, with caveats, that star formation episodes increase galaxy size, as measured by the half-light radius, by about 8\%. In addition to providing a statistically-sized sample with which to explore the star formation behavior of these galaxies, this study also provides a way to select galaxies with specific recent star formation histories for spectroscopic follow-up.
\end{abstract}


\section{Introduction}
\label{sec:intro}

The study of star formation in galaxies is a far too large and complex field to summarize here, even if we confine ourselves to dwarf galaxies \citep[see][and the associated conference proceedings for an introduction to issues related to dwarf galaxy evolution]{haynes}. Instead, we consider how one might explore this topic in a manner that simplifies to one or two questions of interest --- at the expense of ignoring other important aspects. Our current interest is in the internal processes that govern star formation in the simplest possible cases. In other words, if one leaves a galaxy alone --- no interactions, no environmental processes such as ram pressure stripping or tidal harassment, no feedback from a supermassive black hole ---  how does it behave? 

This type of approach, and specifically this question, has motivated the study of dwarf galaxies in the field \citep[for examples across the decades, see][]{gerola,davies,simpson,ferrara,mcquinn,martin}. Although it is difficult to find a sample of galaxies that avoids all of these complications, considering low-mass galaxies mitigates the major coherent internal dynamical drivers (bars and spiral arms, minimizes the likelihood of supermassive black holes (although see \cite{kaviraj} for a counterargument), and, avoids, for the most part, major mergers. Low-mass star-forming systems are primarily found away from overdense areas, thereby also mitigating the complication of environmental effects. Dwarf galaxies found in dense environments then probe the differentiating effects of environments. Studies of field dwarf galaxies have generally concluded that low mass galaxies undergo stochastic and episodic bursts of star formation during their lifetimes \citep[cf.,][]{searle,mcquinn,weisz11,kauffmann14}.  

Such episodes, at least those most frequently observed, are usually intense, can have profound effects on the baryonic distribution of these low-mass galaxies \citep{silk}, and in some cases may eventually lead to the morphological transformation of the galaxy \citep{davies}. The key parameters determining the role these bursts play are the strength, frequency, and duration of the bursts. Bursts, in typical low-redshift dwarf galaxies that are undergoing or have had a recent burst, have star formation rates (SFRs) that are at least a few times the average rate of star formation over the galaxy's lifetime, form between a few and $\sim$ 20\% of the total stellar mass of a system, last about half a Gyr, and have a frequency of roughly one per Gyr \citep{mcquinn}.

Low-surface-brightness (LSB) dwarfs might be considered a contrast to these systems. At least those in the field should have the fuel needed to drive star formation, perhaps even the vigorous bursts identified in other dwarf galaxies, and yet they are diffuse and have low surface brightness, suggesting that at least recently they have not had such a burst.
In particular, the most massive LSB dwarfs, where there is no question that they should be able to gravitationally retain their gas, present a dilemma.  

These are the ultra-diffuse galaxies (UDGs), a term coined by \cite{vanDokkum+2015} upon their identification of a large number of galaxies with low central surface brightness ($\mu_{0,g} > 24$ mag arcsec$^{-2}$) and large  effective radii ($r_e > 1.5$ kpc) in the Coma cluster.  
When, as was the case for \cite{vanDokkum+2015}, these galaxies are found in dense environments, the lack of star formation is attributed to environmental effects \citep{vanDokkum+2015,Carleton+2019,Sales+2020,Tremmel+2020}. When they are found outside of clusters or groups \citep{Prole+2019}, questions arise about what restricts their star formation.

There are various aspects to consider. First, field UDGs/LSBs are more likely to be star-forming (or younger) than their cluster counterparts \citep{Prole+2019,Kadowaki+21}, although there is a quiescent field population \citep{Prole+2021}. As such, perhaps the moderating of star formation is not as absolute as one might have thought necessary. Second, some models for the formation of UDGs invoke strong bursts of star formation, leading to strong stellar feedback, followed by periods of quiescence \citep{DiCintio+2017,Chan+2018}. In such cases, current quiescence does not imply sustained quiescence. In contrast, other models for UDG formation, such as those that invoke high spin halos to halt collapse and subsequent high surface brightness \citep{jimenez,Amorisco+2016,wright}, suggest that star formation may always be modest in these galaxies.

To address these and other topics related to large LSB galaxies, we describe our effort to Systematically Measure Ultra-Diffuse Galaxies (SMUDGes) by  
identifying a sample of UDG candidates across the footprint of the Legacy Survey \citep{Dey+2019} in a series of papers. A strength of this survey is that it probes all galaxy environments. To summarize, the papers in the sequence that are of relevance here are : 1) \cite{Zaritsky+2019}, hereafter Paper I, which presents the survey concept and initial results, 2) \cite{Zaritsky+2021}, hereafter Paper II, which presents the analysis of an equatorial stripe, introduces parameter error estimation, and begins to discuss the effect of Galactic Cirrus on the results, 3) \cite{Zaritsky+2022}, hereafter Paper III, which presents the galaxy catalog obtained for the southern portion of the Legacy Survey that is used here and describes further refinements including, importantly for this study, how redshifts are estimated for those galaxies that can be associated with neighboring galaxies for which redshift measurements exist, 4) \cite{Zaritsky_2023}, hereafter Paper V, which presents the complete galaxy catalog, and 5) \cite{zaritsky+2025}, hereafter Paper VIII, which refines the visual classification of candidates that are unlikely to survive scrutiny. 

It is our aim in this study to use this sample to quantify the recent ($\lesssim 1$ Gyr) star formation history (SFH) of these galaxies and determine the nature of 
SFR fluctuations --- whether the SFR behaves as in other low mass galaxies and whether there is any relation between recent bursts and galaxy size, among other questions. In \S2 we describe our data, including new UV and IR photometry of these sources that is essential to recovering the recent star formation history. In \S3 we describe our process for recovering the SFH of these systems, with a particular focus on the most recent Gyr, where the available diagnostics can best constrain it. We present and discuss our results in \S4 and summarize in \S5.
Throughout, we use a WMAP9 $\Lambda$CDM flat cosmology with $\Omega = 0.287$ and H$_0 = 69.3$ km s$^{-1}$ Mpc$^{-1}$ \citep{hinshaw}. 
Magnitudes are in the AB system \citep{oke1,oke2}. In the appendix, we compare our photometry to that available in the literature and test our ability to recover certain aspects of the SFH of these galaxies. This paper is the first of several underway that will use these results to investigate the governors of  star formation in local, low surface brightness galaxies across different environments.

\section{The Data}

\subsection{The Sample}
\label{sample}
We begin with the 5598 UDG candidates identified in Paper III from our analysis of all images obtained with DECam \citep{decam} at the CTIO 4m, or Blanco, telescope that are included in Data Release 9 (DR9) of the DESI Legacy Imaging Surveys \citep{Dey+2019}. We limit the analysis to this southern portion of the Legacy Survey because of the superior image quality of that section. Following the revisited visual classifications presented in Paper VIII, we reject a few of these candidates and retain 5428. 

We extend the wavelength range of the spectral energy distribution (SED) that we have available to analyze by augmenting the SMUDGes $g$, $r$, and $z$ measurements with ultraviolet bands, FUV (1539 \AA) and NUV (2316 \AA) from the \textit{Galaxy Evolution Explorer} \citep[GALEX;][]{galex}, and infrared bands, W1 (3.4 $\mu$m), W2 (4.6 $\mu$m), W3 (12$\mu$m), and W4 (22 $\mu$m) from the \textit{Wide-field Infrared Survey Explorer} \citep[WISE;][]{Wright_2010}. For our analysis, we require acceptable photometry in all three optical bands and in at least one band in each of the UV and IR regimes (see Appendix \ref{AppSelect} for a discussion of this requirement). This requirement reduces the sample down to 4440 candidates.

GALEX imaging of the sky is limited, and many regions were observed using short exposure times, resulting in poor measurements for these faint galaxies. We conclude that robust photometry  requires a minimum GALEX exposure time of 900 seconds for our sources and therefore add this as a  requirement.  After applying this additional requirement, we have 2428 candidates available for further processing. For these candidates, we provide our non-optical photometric measurements in Table \ref{tab:phot}. Our photometry procedure is described in \S\ref{subsubsec:UV} and \S\ref{subsubsec:photIR}, for the UV and IR bands, respectively.

\begin{deluxetable}{lrrrrrrrrrrrr}
\tablewidth{\columnwidth}
\tablecaption{SMUDGes GALEX UV and WISE IR Photometry}
\label{tab:phot}
\tablehead{
\colhead{SMUDGes ID}&
\colhead{FUV} &
\colhead{$\sigma_{FUV}$}&
\colhead{NUV}&
\colhead{$\sigma_{NUV}$}&
\colhead{W1}&
\colhead{$\sigma_{W1}$}&
\colhead{W2}&
\colhead{$\sigma_{W2}$}&
\colhead{W3}&
\colhead{$\sigma_{W3}$}&
\colhead{W4}&
\colhead{$\sigma_{W4}$}
}
\startdata
SMDG0039472$-$370027&    4.53&    0.17&    4.60&    0.30&    0.00&    2.25&    0.00&    3.07&    0.00&   39.58&  371.88&  321.40\\
SMDG0039489$-$162223&   $-$9.99&   $-$9.99&   $-$9.99&   $-$9.99&   $-$9.99&   $-$9.99&   $-$9.99&   $-$9.99&   $-$9.99&   $-$9.99&   $-$9.99&   $-$9.99\\
SMDG0039525+034347&    0.00&    0.12&    0.00&    0.21&    0.00&    1.62&    0.00&    1.98&    0.00&   40.29&    0.00&  112.52\\
SMDG0039587+024558&    0.31&    0.15&    0.72&    0.23&    0.00&    1.02&    0.00&    1.63&    0.00&   30.02&    0.00&   85.31\\
SMDG0040029+025725&    0.00&    0.13&    0.00&    0.24&    2.71&    2.19&   $-$9.99&   $-$9.99&  113.97&   43.35&    0.00&   61.44\\
SMDG0040103$-$303513&   $-$9.99&   $-$9.99&   $-$9.99&   $-$9.99&   $-$9.99&   $-$9.99&   $-$9.99&   $-$9.99&   $-$9.99&   $-$9.99&   $-$9.99&   $-$9.99\\
SMDG0040194+033200&    1.77&    0.11&    2.70&    0.29&    0.00&    1.95&    0.00&    2.73&   $-$9.99&   $-$9.99&    0.00&  100.21\\
SMDG0040218$-$220213&   $-$9.99&   $-$9.99&   $-$9.99&   $-$9.99&   $-$9.99&   $-$9.99&   $-$9.99&   $-$9.99&   $-$9.99&   $-$9.99&   $-$9.99&   $-$9.99\\
SMDG0040239$-$013230&   $-$9.99&   $-$9.99&    0.37&    0.23&    0.00&    0.85&    0.00&    1.40&    0.00&   22.97&    0.00&   65.61\\
SMDG0040340$-$561554&   $-$9.99&   $-$9.99&   $-$9.99&   $-$9.99&   $-$9.99&   $-$9.99&   $-$9.99&   $-$9.99&   $-$9.99&   $-$9.99&   $-$9.99&   $-$9.99\\
\enddata
\tablecomments{Photometry is presented here for a random set of 10 objects from the SMUDGes catalog \citep{Zaritsky_2023}. Table is line matched to the SMUDGes catalog and the full Table is available electronically. Entries of $-9.99$ indicated no suitable data, while entries with 0.00 indicate no detection and the associated uncertainties provide information on upper flux limits. All fluxes are in units of nanomaggies.}
\end{deluxetable}

Finally, because a measured redshift is necessary to estimate a distance and derive physical parameters, such as the stellar mass, we further limit the sample analyzed here to the subsample of 966 SMUDGes that have an existing redshift estimate. Our SED fitting is unstable if we attempt to simultaneously fit the redshift. Of these 966 galaxies, 159 have spectroscopic measurements from a range of sources that we continually update (see Paper VIII for details and contact authors for an up-to-date list), 701 are projected onto rich clusters with known distances (e.g., Coma, Virgo, and Fornax), and 106 are in groups where other members have known redshifts (see Paper III for details on this subsample). Although cluster associations have proven to be relatively robust \citep[e.g.,][]{Kadowaki+2017,Kadowaki+21}, in Paper III we estimate that up to $\sim$ 30\% of redshifts based on group membership may be catastrophically wrong. This error rate implies that perhaps 32 galaxies in the sample have such redshift errors. We will show that such errors do not qualitatively affect our conclusions in this study. 

The redshift requirement introduces certain biases in the sample. First, the sample tends to be biased toward dense environments because many of the redshift estimates are for galaxies associated with dense environments. Among the field galaxies in the sample, we tend to be biased toward star forming systems because it is easier to measure a redshift from an optical spectrum if emission lines are present and because a significant number of our field redshifts come from H{\small I} observations. As such, one should be cautious when drawing inferences from the relative number of galaxies of certain types. We will analyze our star formation history results in the context of  different environmental measures in Khim et al. (in prep.) and present those environmental measures in Narisetty et al. (in prep.).

As in our previous studies, all processing and analysis is performed using the Puma cluster at the University of Arizona High Performance Computing center\footnote{\href{https://public.confluence.arizona.edu/display/UAHPC/Resources}{public.confluence.arizona.edu/display/UAHPC/Resources}}.

\subsection{Optical Photometry}

Our image processing pipeline for the DECam optical bands is described in detail in Papers I, II, III, and V. We summarize only here.  Briefly, for the Southern Survey (Papers I, II, and III), we obtained calibrated CCD images and data quality masks that have been processed by the DECam Community Pipeline \citep{valdes} from the NOIRLab website. 
Our approach to object detection is primarily described in Paper I. We start by masking objects that are too bright to be UDGs and then use tailored filters to select targets that may match our size requirements ($r_e \gtrsim$ 5 arcsec). 
We then measure the photometric and structural properties for each detection using GALFIT \citep{Peng+2002} and retain viable candidates.  Detections likely to be cirrus are culled using $\tau_{353}$ \citep{planck} and WISE 12 $\mu$m \citep{meisner} dust maps.  Specifically, we  reject those at locations  with single-point values greater than 0.05 for $\tau_{353}$ or 0.1 MJy/sr for WISE 12 $\mu$m. Further screening is performed with an automated system using a convolutional neural network previously trained with known candidate UDGs.  Each of the surviving candidates is then visually reviewed. The four structural parameters derived from the optical GALFIT estimates (effective radius, $r_e$; axis ratio, $b/a$; S\'ersic index, $n$; and position angle, $\theta$) are used to measure the magnitudes in the UV and IR bands.

Our approach for estimating the uncertainties of the optical measurements is described in detail in Paper II with modifications noted in Papers III and V.  We use the values obtained from Paper V.  Briefly, we estimate uncertainties by planting simulated UDGs using S\'ersic profiles with random structural and photometric properties at random locations.  These were processed separately with the same pipeline used for our real sources, including automated classification. We define errors for these simulations as the difference between their final GALFIT values and the values used when creating them (GALFIT $-$ input). These uncertainties are generally asymmetric, and we estimate the ``1$\sigma$" confidence limits as the 16 and 84 percentiles of the distribution for a set of similar simulated objects. Finally, we model the simulated uncertainties using a second-order polynomial to fit the simulation data and apply these results to the optical data.  

\subsection{Ultraviolet Photometry}
\label{subsubsec:UV}
UV images of galaxies tend to be clumpy and, in general, are not well represented by S\'ersic profiles.  Therefore, we measure the magnitudes and associated uncertainties using aperture photometry.  All GALEX images were obtained from the Mikulski Archive for Space Telescopes (MAST)\footnote{\href{http://galex.stsci.edu/data/}{http://galex.stsci.edu/data/}}\footnote{\dataset[galex]{galex}} and can be accessed via \dataset[https://doi.org/10.17909/npwd-sh81]{https://doi.org/10.17909/npwd-sh81} \citep{galexDOI}. Downloaded images include the raw number of counts per pixel (cnt), the high resolution relative response (rrhr, which is the relative sensitivity multiplied by the exposure time map), and the sky background map (skybg).  For each candidate, we download all observations in which the candidate's centroid lies less than 0.55 degrees from the center of the field of view. This approach rejects data where the target is in a part of the field-of-view where the image quality deteriorates. We then stack the available images for each candidate. After creating cutouts centered on the candidate, an intensity image is generated as cnt/rrhr.  Native GALEX images are not background-subtracted so skybg is subtracted from the intensity image.  Using the structural parameters obtained in the optical, an effective centered aperture of 1.5$r_e$ is calculated by convolving the associated filter point-spread function  \citep[FWHM in FUV of 4.2\arcsec and in NUV of 5.2\arcsec;][]{galex} with the object's S\'ersic profile and then matching the flux within that aperture in the convolved profile to the flux at 1.5$r_e$ in the original profile. 

Although we subtract the sky level from the intensity image, the aperture can still be contaminated by projected objects. To estimate both the level of contamination and the associated uncertainties, we define 24 wedge-shaped apertures in two annuli surrounding the target starting at a distance of 3$r_e$ from its center.
This step is done using the observation with the longest exposure time, which is generally roughly consistent with the total exposure time (on average the longest exposure comprises 87\% of the total exposure time). Each annulus is configured such that its total area is 12 times that of the 1.5$r_e$ aperture used for the target.  Small differences in area between wedges  and the target aperture are corrected by multiplying the flux within a wedge by the ratio of the target aperture area to that of the wedge.  In general, the distributions of enclosed flux within the background apertures are not Gaussian, and, therefore, we estimate the contamination level and the 1$\sigma$ confidence limits using the Python astropy.stats functions biweight\_location and biweight\_scale, respectively \citep{beers}. We generate our final estimate for the target GALEX flux by subtracting the contamination level and multiplying by the aperture correction factor. 
We adopt the FUV and NUV zeropoints of 18.82 and 20.08 AB mag \citep{galex}. Any resulting flux value less than zero has its associated magnitude flagged. 

Although the above describes our basic approach, 
there are additional issues that we address to ensure that the results are reliable.  First, although the targets are chosen to be 0.55 deg away from an image edge, the background apertures may be closer to an image edge and, therefore, unreliable. In such cases, the number of available background wedges may be significantly reduced, thereby reducing the robustness of our statistics.  Moreover, if a bright UV object lies near a target or background aperture, it will skew the measurement.  We mitigate this potential contamination using the $g-$band Legacy Survey images\footnote{\href{https://www.legacysurvey.org/viewer/}{www.legacysurvey.org/viewer}}.
Each UV aperture is evaluated separately. If there is an object that is too bright to be a UDG ($\mu_{0,g} < 24$ mag arcsec$^{-2}$) within 10$^{\prime\prime}$ of its aperture edge, then that object is masked out to the $g-$band 26.5 mag arcsec$^{-2}$ isophote. 
Background apertures that have more than 20\% of their pixels masked are discarded. Masked pixels are converted to NaN's and then 
replaced with interpolated values using the numpy function interpolate\_replace\_nans. We found that this gave poor results when the g-band object flux was $>15$ nanomaggies. Therefore, background wedges containing objects that exceed this value are discarded.  Results obtained using fewer than 5 background wedges or where the target aperture itself was discarded are flagged as unacceptable.  A total of 117 UV observations are rejected by these criteria, with 42 failing in both bands. In these cases, the photometry is set to $-$9.99 in Table \ref{tab:phot}, the same value as if there are no available data.

\subsection{Infrared Photometry}
\label{subsubsec:photIR}
We use a modified version of forced photometry for the infrared bands, fixing the structural parameters to be those derived from the optical data but unfixing the centroid. The centroid freedom allows for slight astrometric differences. We download the WISE images as coadded bricks from Data Release 10 of the southern portion of the Legacy Survey\footnote{\href{https://portal.nersc.gov/cfs/cosmo/data/legacysurvey/dr10/south/coadd}{portal.nersc.gov/cfs/cosmo/data/legacysurvey/dr10/south/coadd}}.  As with the optical pipeline used in Paper V, both the science and inverse-variance images are used in the GALFIT processing. For each target, we create cutouts of 325$^{\prime\prime}$ on a side for further processing.  Because the WISE pixel scale is 2.75$^{\prime\prime}$ pix$^{-1}$, the cutouts are first resampled to 0.262$^{\prime\prime}$ pix$^{-1}$ to match the optical images. We then create a mask using SEP \citep{sep} with a threshold set at the background rms level. 
After three sigma-clipping, we estimate the median sky level and its uncertainty. As with the optical images, we use wavelets to isolate objects at different
scales with a tailored filter (see Paper I for details).  However, while wavelets were used to detect potential candidates in the optical, here they are used to mask point sources that may affect GALFIT results in crowded IR fields. We measure the apparent magnitudes using GALFIT and the optical structural parameters. We flag the results if GALFIT failed to converge or if the estimates were found to be unreliable. Although these requirements led to a total of 1576 IR observations being flagged, none were rejected in all four bands.  WISE stacked images are on the Vega system and are converted to AB magnitudes using \begin{equation}
\label{eq:vega2ab}
    mag_{AB} = mag_{Vega} + \Delta mag
\end{equation}
where $\Delta mag =$ 2.699, 3.339, 5.174, and 6.620 mag, for W1 through W4, respectively\footnote{\href{https://wise2.ipac.caltech.edu/docs/release/allsky/expsup/sec4_4h.html\#conv2ab}{wise2.ipac.caltech.edu/docs/release/allsky/expsup/\-sec4\_4h.html\#conv2ab}}.

\begin{figure}[ht]
\begin{center}
\includegraphics[width=0.47\textwidth]{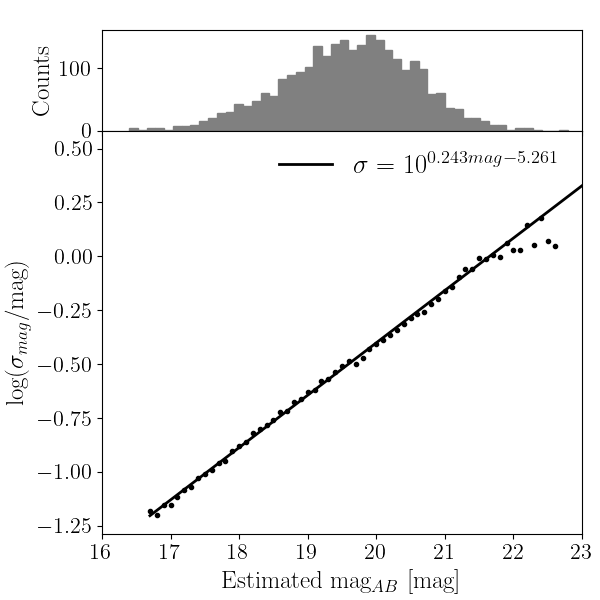}
\end{center}
\vskip -.5cm
\caption{The WISE W1 photometric 1$\sigma$ uncertainties determined assuming that the error distributions derived from simulations are Gaussian (bottom panel).
Histogram (top panel) represents the W1 magnitude distribution of the science targets. The fitted functional form is given (see Table \ref{tab:WISE} for corresponding coefficients for the other bands).}
\label{fig:W1errors}
\end{figure}

The uncertainty estimates produced by GALFIT have been shown to significantly underestimate true errors \citep{Haussler}. As with our optical measurements, we estimate the uncertainties using simulations. Due 
to the low completeness at faint magnitudes, the number of simulated sources required for statistical robustness
ranged from about 10$^6$ for W4 to $5\times10^6$ for W3. We divide the simulated sources by magnitude into 0.1 mag bins and fit the resulting measured parameter histogram with a Gaussian.  We find that for faint objects, GALFIT sometimes moves the object centroid to a nearby object rather than failing.  We mitigate this problem by requiring that the offset between the fitted GALFIT and optical locations be $<0.4r_e$ for both science and simulated targets.  As we show in Figure \ref{fig:W1errors}, until sky noise begins to dominate the results (at about 21.5 mag in this example), there is a linear correlation between the logarithm of uncertainty, $\log(\sigma_{mag})$, and magnitude. We use these results to estimate uncertainties using
\begin{equation}
    \sigma_{mag} = 10^{a*mag +b},
\end{equation}
where a and b are the band-dependent slopes and intercepts given in Table \ref{tab:WISE}.  

\begin{deluxetable}{lrr}
\tablewidth{\columnwidth}
\tablecaption{WISE Uncertainty Fit Coefficients} 
\label{tab:WISE}
\tablehead{
\colhead{Band    }&
\colhead{a} &
\colhead{b} }
\startdata
W1 & 0.243 & $-$5.261 \\
W2 & 0.295 & $-$6.152 \\
W3 & 0.374 & $-$6.876 \\
W4 & 0.424 & $-$6.843 \\
\enddata

\end{deluxetable}

\subsection{The Photometric Catalog}

We provide UV and IR photometry in Table \ref{tab:phot} for 2428 SMUDGes. The Table is line matched to the Paper V SMUDGes catalog, so many lines contain no photometry (all entries are $-9.99$). 
For the UV bands, if the magnitude is flagged, indicating that its associated flux was $\leq 0$, we assigned a flux value of zero.
Similarly, we assign a UV flux value of zero if the flux is less than the 1$\sigma$ upper bound. For the IR bands, if the flux is less than the uncertainty or if the offset between the GALFIT IR and optical centers is $>0.4r_e$, then the flux is set to zero. For these cases, we conservatively set the uncertainty to be the flux contained within the corresponding S\'ersic profile that has a peak flux equal to the noise in the sky. 

\section{Deriving the Star Formation History}
\label{sec:SED}
We model the SEDs  using PROSPECTOR \citep{prospector}, a Bayesian forward-modeling code developed to infer stellar population parameters from galaxy photometry.  This software uses the publicly available Flexible Stellar Population Synthesis (FSPS) code \citep{conroy2009, conroy2010}.
We now proceed to provide details on our application of PROSPECTOR in SMUDGes.

\subsection{Curating the Input}
\label{curating}

Some pre-processing of the data is needed before running PROSPECTOR. 
First, we correct apparent magnitudes for Galactic extinction using dustmaps.py \citep{green} to obtain $E$($B$-$V$)$_{SFD}$, where SFD references the extinction maps of  \cite{SFD}. We obtain the optical and WISE extinction coefficients ($A_g=3.214, A_r=2.165, A_z=1.211, A_{W1}=0.184, A_{W2}=0.113, A_{W3}=0.024, A_{W4}=0.009$) from the Legacy Survey\footnote{\href{https://www.legacysurvey.org/dr10/catalogs/\#galactic-extinction-coefficients}{www.legacysurvey.org/dr10/catalogs/$\#$galactic-extinction-coefficients}} and  those for GALEX ($A_{FUV}=8.01, A_{NUV}=6.79$) from \cite{wall}. First, we correct AB magnitudes, $mag_{\rm AB}$ by subtracting $A_X \times E(B-V)_{SFD}$ from the uncorrected values where $A_X$ is the associated extinction coefficient for the relevant filter band.  Second, we convert from $mag_{\rm AB}$ to flux
in maggies using
\begin{equation}
\label{eq:mag2flux}
    flux = 10^{-9}\bigl(10^{(22.5-mag_{\rm AB})/2.5}\bigr)
\end{equation}
where 22.5 is the zeropoint for the conversion. Similarly, we  convert the magnitude uncertainties for optical and WISE images, $\sigma_{mag_{\rm AB}}$, to the flux uncertainties, $\sigma_{flux}$, using
\begin{equation}
    \sigma_{flux} = \frac{ln(10)}{2.5}\sigma_{mag_{\rm AB}}.
\end{equation}
For the optical, magnitudes and their uncertainties are obtained from the SMUDGes catalog available in Paper III. 
PROSPECTOR expects symmetric uncertainties, so we average the upper and lower errors provided in the catalog. 
%

\subsection{Recovering The Star Formation History}
\label{sec:SFH}

PROSPECTOR allows one to fit either a  parametric SFH model (an SFR with a pre-determined functional form across time) or a nonparametric one, where the SFR is (relatively) freely determined within separate age bins.  We opt to fit a nonparametric SFH because we are focusing on unknown variations in the recent SFR. This choice allows maximum flexibility in the SFH in a limited set of time bins \citep{prospector, Leja_2019}. 

An important choice lies in the definition of the age bins.  We set our age bins to roughly match the ages where the different diagnostics available to us differentiate best. We set four age bins ($<20$ Myr, 20-150 Myr, 150 Myr-1 Gyr, and $>1$ Gyr).  
The first age bin is best tracked by H$\alpha$ emission and tracks ongoing star formation. While we do not currently have H$\alpha$ imaging broadly for SMUDGes, there are subsamples \citep[e.g.,][]{loraine} for which H$\alpha$ spectroscopy is beginning to become available. The second bin is best probed by the GALEX UV flux and indicates recent or ongoing star formation. While the UV fades after that, the optical (blue colors, Balmer lines) tracks star formation within the last $\sim$ Gyr and so this intermediate population is measured in the third bin. Finally, at ages $>$ 1 Gyr, we are primarily tracking the integrated stellar mass, and we have few diagnostics from broad band photometry with which to recover star formation rate variations. IR photometry  provides the most robust measure of the total stellar mass. 

There is, however, a difficult choice in the degree of freedom that one is willing to allow in bin-to-bin SFR variations. It is likely unphysical to allow for arbitrarily large variations because the SFR  is unlikely to change rapidly in correspondence to arbitrary bin boundaries or to be tightly coordinated across an entire galaxy.  PROSPECTOR offers several options for the adopted priors to help the user control the bin-to-bin SFR variations. The most common ones cited are one that is termed the continuity prior \citep[cf.,][]{Leja_2019, suess, haskell, das, Nersesian, Turner, Williams, Reyes} and another that is termed the Dirichlet prior \citep[cf.,][]{Leja_2019, Lower, Ji, Williams, Tanaka, haskell, Webb, Reyes}. The continuity prior fits for $\Delta$log(SFR) between adjacent time bins \citep{Leja_2019} using a prior that favors small values of $\Delta$log(SFR), disfavoring sharp transitions in star formation rates. Although this dampens sharp changes between adjacent age bins, it is considered sufficiently flexible to handle both bursts and quenching \citep{Leja_2019, suess, haskell, das}. The Dirichlet prior models the fractional specific SFR in each time bin using a Dirichlet distribution \citep{Leja_2019}.  The Dirichlet prior as used in PROSPECTOR contains a concentration parameter $\alpha_D$, such that high values ($\gtrsim$1) tend to smooth the distribution between bins, while lower values prefer to concentrate the mass in single bins with results strongly dependent on the selected value \citep{Leja_2019}.

We investigate this choice in more detail in Appendix \ref{app:current_sfr} using simulations and a wide range of $\alpha_D$ selections. To summarize, 
because our population consists of galaxies with diverse environments, colors, and physical characteristics, we expect large variations in global SFHs and, therefore, choose to use the more general continuity prior.  Our decision is further reinforced by the work of \cite{Reyes} who note that PROSPECTOR has trouble fitting low-mass galaxies using the Dirichlet prior below some minimum number of age bins and for certain $\alpha_D$ values.  In particular, they suggest using at least six age bins and restricting $\alpha_D$ to values $\gtrsim$0.7. We confirm their findings in Appendix \ref{app:current_sfr}.

Next, we outline some more technical choices.   First,
we choose to set the continuity nonparametric prior to the default Student’s t-distribution.  Second, we use the Kroupa initial mass function \citep{kroupa}, the MILES spectral library \citep{miles} and the MIST isochrones \citep{mist1, mist2, mist3, mist4, mist5}. 
Third, low-mass UDGs tend to be metal-poor \citep{Ruiz-Lara+2018,Ferre-Mateu+2018,barbosa,heesters} and, therefore, we adopt the SMC bar extinction law from \cite{gordon2003}. 
We adopt an initial dust prior that is uniform from $-$0.4 to 2 mag and ignore dust emission\footnote{Dust emission is potentially a factor only in the W3 and W4 bands \citep{remy}, where we mostly have upper limits. The small number of galaxies, 8(4), for which we have a W3(W4) measurement that exceeds a non-detection by $2\sigma$ is entirely consistent with attribution to random noise.}. The unphysical lower limit allows for full exploration of this parameter given the uncertainties and degeneracies between the parameters. We discuss this choice in more detail in Appendix \ref{App:Dust}.
Fourth, we fit for the log of the total stellar mass (\textit{logmass}) in solar units and metallicity (\textit{logzsol}). 
We adopt uniform priors of 2.5 - 10  for \textit{logmass} and $-$4 to 0.2 dex for \textit{logzsol}.  Bayesian posteriors are estimated using the \textit{dynesty} package \citep{speagle, koposov} based on a nested sampling algorithm described by \cite{higson} and \cite{skilling2004,skilling2006} with 200 live points.

We estimate the SFR and the total mass for each age bin using the PROSPECTOR functions logsfr\_ratios\_to\_sfrs and logsfr\_ratios\_to\_masses.   We define the lower and upper uncertainties as the 16\textsuperscript{th} and 84\textsuperscript{th} percentiles, respectively.

\section{Results}

\begin{deluxetable*}{lr}
\caption{Entry Description for the Star Formation History SMUDGes Table$^a$}
\label{tab:sfh}
\tablehead{
\colhead{Column Name}&
\colhead{Description}\\
}
\startdata
SMDG& SMUDGes Catalog Name\\
logzsol & [Fe/H] \\
logzsol-upper& 1$\sigma$ upper bound on [Fe/H]\\
logzsol-lower& 1$\sigma$ lower bound on [Fe/H] \\
A$_V$ & internal V-band$^b$ extinction [mag]\\
A$_{V}$-upper&1$\sigma$ upper bound on internal V-band extinction [mag]\\
A$_{V}$-lower&1$\sigma$ lower bound on internal V-band extinction [mag]\\
M$_*$&log10(M$_*$/M$_\odot$)\\
M$_{*}$-upper&1$\sigma$ upper bound on log10(M$_*$/M$_\odot$)\\
M$_{*}$-lower&1$\sigma$ lower bound on log10(M$_*$/M$_\odot$)\\
SFR$_0$&log10(current star formation rate/(M$_*$ yr$^{-1}$))\\
SFR$_{0}$-upper&1$\sigma$ upper bound on log10(SFR$_0$/(M$_*$ yr$^{-1}$))\\
SFR$_{0}$-lower&1$\sigma$ lower bound on log10(SFR$_0$/(M$_*$ yr$^{-1}$))\\
M$_{*,0}$&log10(stellar mass in most recent age bin/M$_\odot$)\\
M$_{*,0}$-upper&1$\sigma$ upper bound on log10(M$_{*,0}$/M$_\odot$)\\
M$_{*,0}$-lower&1$\sigma$ lower bound on log10(M$_{*,0}$/M$_\odot$)\\
M$_{*,1}$&log10(M$_{*,1}$/M$_\odot$)\\
M$_{*,1}$-upper&1$\sigma$ upper bound on log10(M$_{*,1}$/M$_\odot$)\\
M$_{*,1}$-lower&1$\sigma$ lower bound on log10(M$_{*,1}$/M$_\odot$)\\
M$_{*,2}$&log10(M$_{*,2}$/M$_\odot$)\\
M$_{*,2}$-upper&1$\sigma$ upper bound on log10(M$_{*,2}$/M$_\odot$)\\
M$_{*,2}$-lower&1$\sigma$ lower bound on log10(M$_{*,2}$/M$_\odot$)\\
M$_{*,3}$&log10(M$_{*,3}$/M$_\odot$)\\
M$_{*,3}$-upper&1$\sigma$ upper bound on log10(M$_{*,3}$/M$_\odot$)\\
M$_{*,3}$-lower&1$\sigma$ lower bound on log10(M$_{*,3}$/M$_\odot$)\\
sSFR$_0$& log10(specific star formation rate in most recent age bin/yr$^{-1}$)\\
sSFR$_{0}$-upper&1$\sigma$ upper bound on log10(sSFR$_0$/yr$^{-1}$)\\
sSFR$_{0}$-lower&1$\sigma$ lower bound on log10(sSFR$_0$/yr$^{-1}$)\\
sSFR$_1$&log10(specific star formation rate in second age bin/yr$^{-1}$)\\
sSFR$_{1}$-upper&1$\sigma$ upper bound on log10(sSFR$_1$/yr$^{-1}$)\\
sSFR$_{1}$-lower&1$\sigma$lower bound on log10(sSFR$_1$/yr$^{-1}$)\\
sSFR$_2$&log10(specific star formation rate in third age bin/yr$^{-1}$)\\
sSFR$_{2}$-upper&1$\sigma$ upper bound on log10(sSFR$_2$/yr$^{-1}$)\\
sSFR$_{2}$-lower&1$\sigma$ lower bound on log10(sSFR$_2$/yr$^{-1}$)\\
sSFR$_3$& log10(specific star formation rate in fourth age bin/yr$^{-1}$)\\
sSFR$_{3}$-upper&1$\sigma$ upper bound on log10(sSFR$_3$/yr$^{-1}$)\\
sSFR$_{3}$-lower&1$\sigma$ lower bound on log10(sSFR$_3$/yr$^{-1}$)\\
f-SFH&analysis flag (2 = has all non-optical photometric data (2 UV + 4 IR), \\
&1 = meets minimal necessary photometry requirement, 0 = missing necessary redshift or photometry)\\
\enddata
\tablenote{The complete catalog is available as the electronic version of this Table.}
\tablenote{Extinction values are allowed to be negative to avoid biasing other parameters (see text). PROSPECTOR returns values for the optical depth. We correct to magnitudes of extinction by multiplying by 1.086 and present those values here.}
\end{deluxetable*}

We present the PROSPECTOR results in Table \ref{tab:sfh} for the 966 SMUDGes with distance estimates that are used in all of the following analysis. In our presentation, we use certain combinations of the results to provide intuition and synthesize this rich dataset. In particular, we will discuss the current star formation as measured by the mass of stars in the first bin, denoted by subscript 0, divided by the time duration of that bin (SFR$_0$ $\equiv M_0/20$ Myr), the current specific SFR, which will be SFR$_0$ divided by the total stellar mass (sSFR$_0 \equiv {\rm SFR_0}/M_*$), and the intermediate age stellar mass fraction, which we define to be the stellar mass in the 3rd bin divided by the 
stellar mass in the 4th bin ($\equiv M_2/M_3$, hereafter referred to as A/K, reflecting the ratio of A-type stars to K-type stars for historical reasons and as shorthand).
We have opted not to include $M_1$ in the intermediate age bin to provide greater statistical independence between A/K and the SFR. The M$_1$ bin is likely to be dominated by B-type stars if there has been recent star formation. 

A key consideration in interpreting the results is the ``leakage" of inferred star formation from one temporal bin to the next resulting both from photometric uncertainties and from our PROSPECTOR priors. We have previously discussed why smoother star formation histories rather than ones with abrupt changes that correspond to the edges of the defined bins may be appropriate, nevertheless certain scenarios are disfavored due to this choice. 
Because of this challenge, results and the corresponding inferences must be vetted with either spectroscopic observations, which offer more temporal fidelity, or simulations, where one can test whether a specific scenario does indeed produce the anticipated signature in the PROSPECTOR results. 
The reader should bear this in mind during our discussion and in any independent application of the star formation rates presented in Table \ref{tab:sfh}. We discuss the effects of leakage on our specific interpretations in Appendices \ref{app:priors} and \ref{app:sims}.

\subsection{Comparison to Published Classifications}

The preferred way to quantify the stellar populations of galaxies is through spectroscopy. There, one can identify emission lines that testify to ongoing star formation (or possible nuclear activity), Balmer absorption lines that highlight intermediate age stars, a Solar-type spectrum that shows the presence of long-lived, lower mass stars, and other diagnostics that can be used to measure metallicity and dust extinction. 
However, even with spectroscopy there are potential pitfalls. Perhaps the most dangerous is that of aperture bias, where the spectral aperture does not welcome light from the entire galaxy. In that case, knots of star formation or a radial gradient may skew the inference. The antidote for this problem is integral field spectroscopy (IFS), where most or all of the galaxy is included in the aperture. IFS observations are presented for a modest number of SMUDGEs by  \cite{loraine} and for other LSBs/UDGs by \cite{Ferre-Mateu+2018}, who happened to include a few SMUDGes within their larger sample. These provide the best comparison set to our results, which are photometric rather than spectroscopic, but which are at least based on the light from the entire galaxy. 

In Figure \ref{fig:evol} we show where the SMUDGes lie within our preferred diagnostic plot of SFR$_0$ vs. A/K. SFR$_0$ provides a measure of the ongoing star formation, so we expect the star forming systems to lie near the top of the Figure. A/K measures the ratio of intermediate age stars (with ages $<$ 1 Gyr) to older stars, representing a measure of the relative strength of the recent star formation rate of the galaxy. Exclusively old stellar populations will lie toward the left of the Figure, and also toward the bottom because they have no ongoing star formation. Those galaxies that may have had recent star formation, but have no current star formation, will lie toward the lower right of the Figure. The degree to which star forming, post-starforming, and quenched galaxies are segregated in this Figure will determine the precision with which we can use it for diagnostic purposes.

We test these expectations by highlighting those SMUDGes in the Figure
with previously published designations using larger plotting symbols. We clarify that the values of SFR$_0$ and A/K are those we measure for these galaxies, not from the literature. The star forming (SF) systems of \cite{loraine} all correctly lie at the upper end of the inferred SFR$_0$ values. This agreement validates the description of this population and confirms that our SED analysis correctly classifies these systems. Galaxies classified as post-starburst (PSB) or quenched by \cite{loraine} and those classified as quenched by \cite{Ferre-Mateu+2018} also fall as expected. The slight overlap between PSBs and quenched galaxies may hint at either some level of uncertainty in our classification or at a problem with the previous classification. The latter could arise if the IFS did not actually cover the full galaxy, which we know to be the case for at least some of the KCWI observations of \cite{loraine}. Despite this partial overlap, the PSB galaxies generally have small uncertainties in A/K, in contrast to the quenched galaxies, which have large uncertainties, and they are segregated in the Figure.

We also compare our results to those obtained for  a sample of galaxies using a photometric SED analysis that is similar to ours  \citep{Buzzo_2022}. The galaxies they classified as quenched  fall along the quenched sequence highlighted by the \cite{Ferre-Mateu+2018} galaxies. These galaxies have very few stars with age $<$ 1 Gyr, so the uncertainties in our log-log plot are large along both axes, but still clearly distinguishable from the star forming systems, and to a lesser degree from the PSBs. The relative coherence of the quenched sample given the large uncertainties suggests that our uncertainty estimates along each axis for the quenched population are not independent. We will return to this topic later.

\medskip
{\sl Conclusion 1: Our SED analysis is capable of identifying SF, PSB, and quenched galaxies at a level of precision comparable to that achieved using integral field spectroscopy.}
\medskip

\subsection{Initial Impressions
}
\begin{figure}[ht]
\begin{center}
\includegraphics[width=0.47\textwidth]{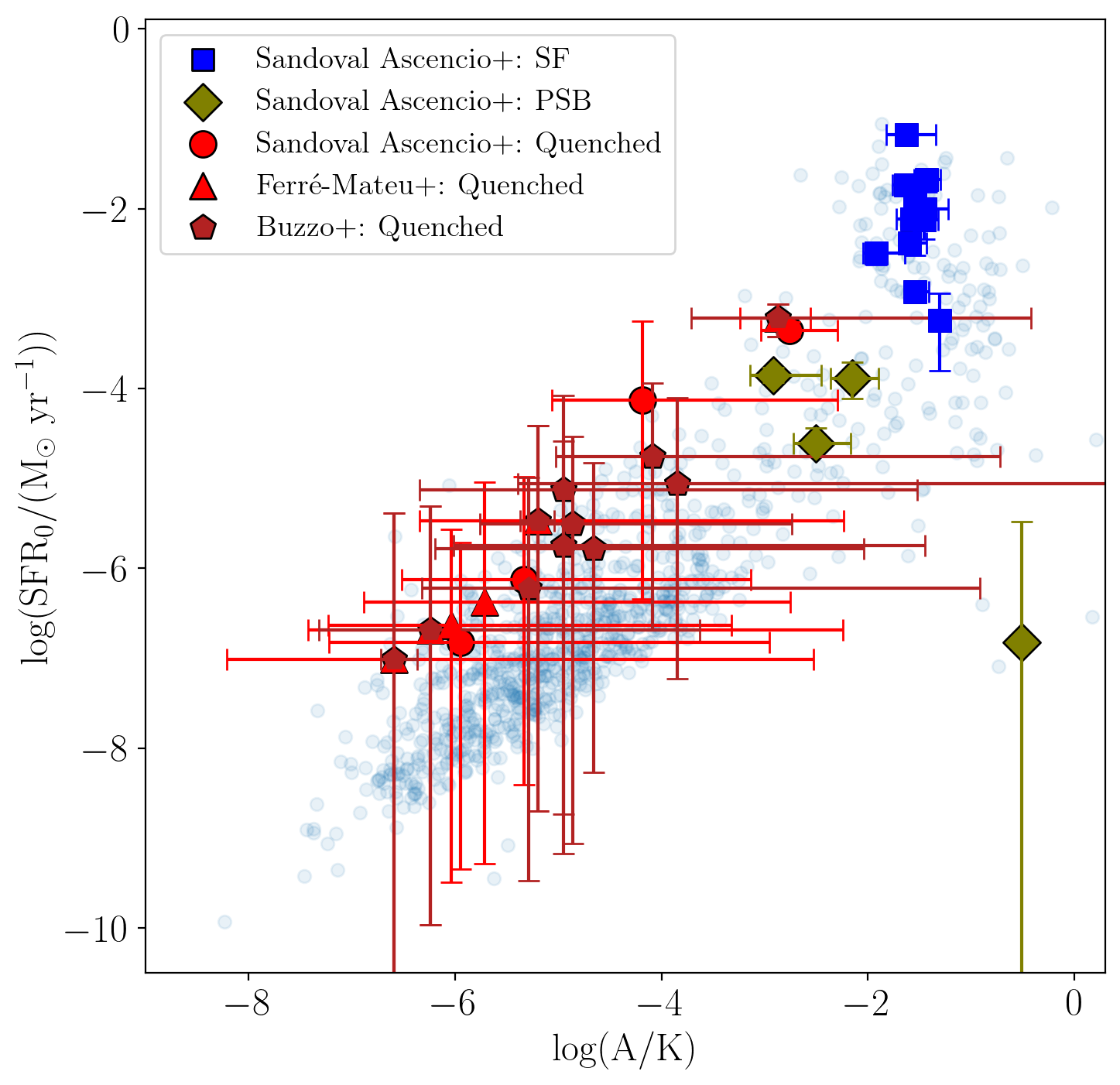}
\end{center}
\vskip -.5cm
\caption{A/K plotted against star formation rate, SFR$_0$, as measured from our PROSPECTOR modeling for SMUDGes galaxies. The larger, filled points with errorbars highlight galaxies for which we have independent classifications from spectroscopic observations \citep{Ferre-Mateu+2018,loraine} or SED analysis \citep{Buzzo_2022}. References for the  independent classifications are given in the legend.}
\label{fig:evol}
\end{figure}

The SMUDGes sample includes LSB dwarf galaxies and UDGs in all environments. The distribution in Figure \ref{fig:evol} has various interesting {\sl apparent} characteristics. First, there is a population of galaxies that has both strong star formation and a relatively high fraction of intermediate age stars (large A/K). We infer that 1) there is a significant population of star forming galaxies in SMUDGes and they can be identified by our analysis, 2) if star formation is episodic, those episodes have coherence over at least $\sim$ Gyr because, for the most part, SFR$_0$ and A/K correlate, and 3) if star formation is episodic, the breaks in star formation activity cannot be $\gtrsim$
1 Gyr because we do not find old systems (low A/K) with ongoing star formation. These first impressions suggest that this diagram can be used to put limits on the timescale of star formation in the low surface brightness galaxies 
that are currently forming stars. We have added the caveat {\sl apparent} because we have not yet established the degree to which the non-parametric modeling faithfully models the actual SFH of these systems. We will return to this question later and find that all is not necessarily as it first appears. In particular, as we discuss in detail in \S\ref{sec:recentSF} and the Appendix, our models are unreliable in identifying galaxies that have been quenched over at least the last Gyr but have recently ($<$ 20 Myr) started forming stars. The models are otherwise broadly reliable (see Appendix).

\begin{figure}[ht]
\begin{center}
\includegraphics[width=0.47\textwidth]{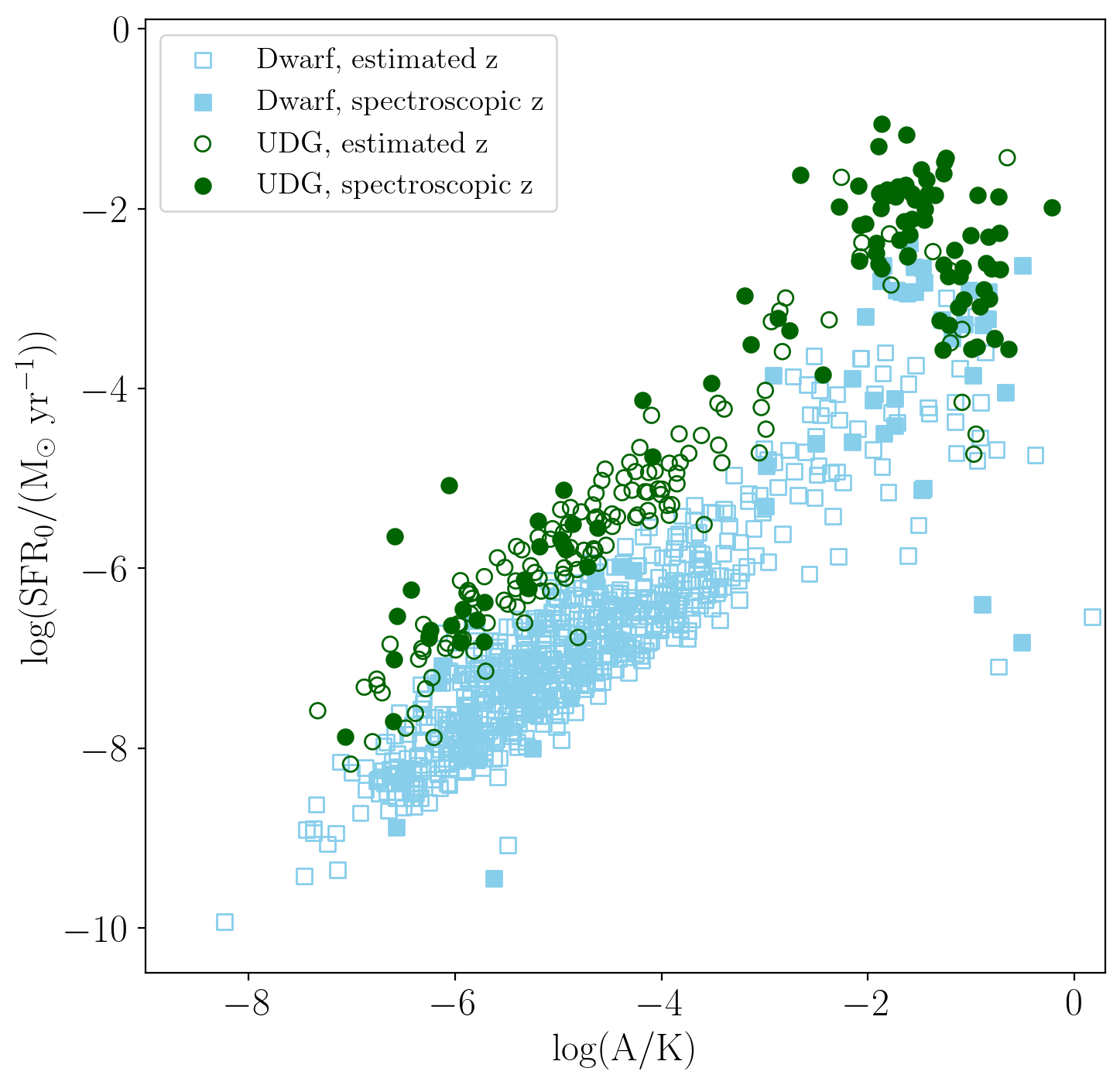}
\end{center}
\vskip -.5cm
\caption{Testing subsamples. We compare the distribution of LSB dwarf galaxies to UDGs and of those with estimated vs. spectroscopic redshifts (see legend). There is a clear offset in SFR$_0$ between LSB dwarfs and UDGs that is presumably due to differences in overall mass when we divide the sample by $r_e$. There is no clear distinction between those galaxies with and without spectroscopic redshifts.}
\label{fig:samples}
\end{figure}

Before proceeding in interpreting these results, we show in Figure \ref{fig:samples} how the results vary both as a function of the redshift source (spectroscopic vs. estimated) and the galaxy type (LSB dwarf ($r_e < 1.5$ kpc) vs. UDG ($r_e \ge 1.5$ kpc)). We see a distinction in SFR$_0$ between the LSB dwarf and UDG populations. This presumably reflects a difference in mass between these two populations, where galaxies of greater mass can also support a higher SFR$_0$. 
We will test this assumption below. We see no evident distinction in the behavior of galaxies with spectroscopic vs. estimated redshifts. As such, we continue to utilize the full sample.

\subsection{Current Star Formation}

We now return to the features of the distribution shown in Figure \ref{fig:evol} that we refer to as apparent. The first of those addressed the population of star forming galaxies. Using simulations (Appendix \ref{app:sims}), we demonstrate that our recovered SFR$_0$ is robust under various conditions, although perhaps with a modest bias of $< 0.4$ dex (Table \ref{tab:SFRrecovery} in the Appendix) relative to the input values in our simulations. Therefore, we are confident to this level of accuracy in the calculated SFR$_0$. For context, SFR$_0$ in the Figure spans 9 dex in comparison to this possible bias of $\le 0.4$ dex.

\begin{figure}[ht]
\begin{center}
\includegraphics[width=0.47\textwidth]{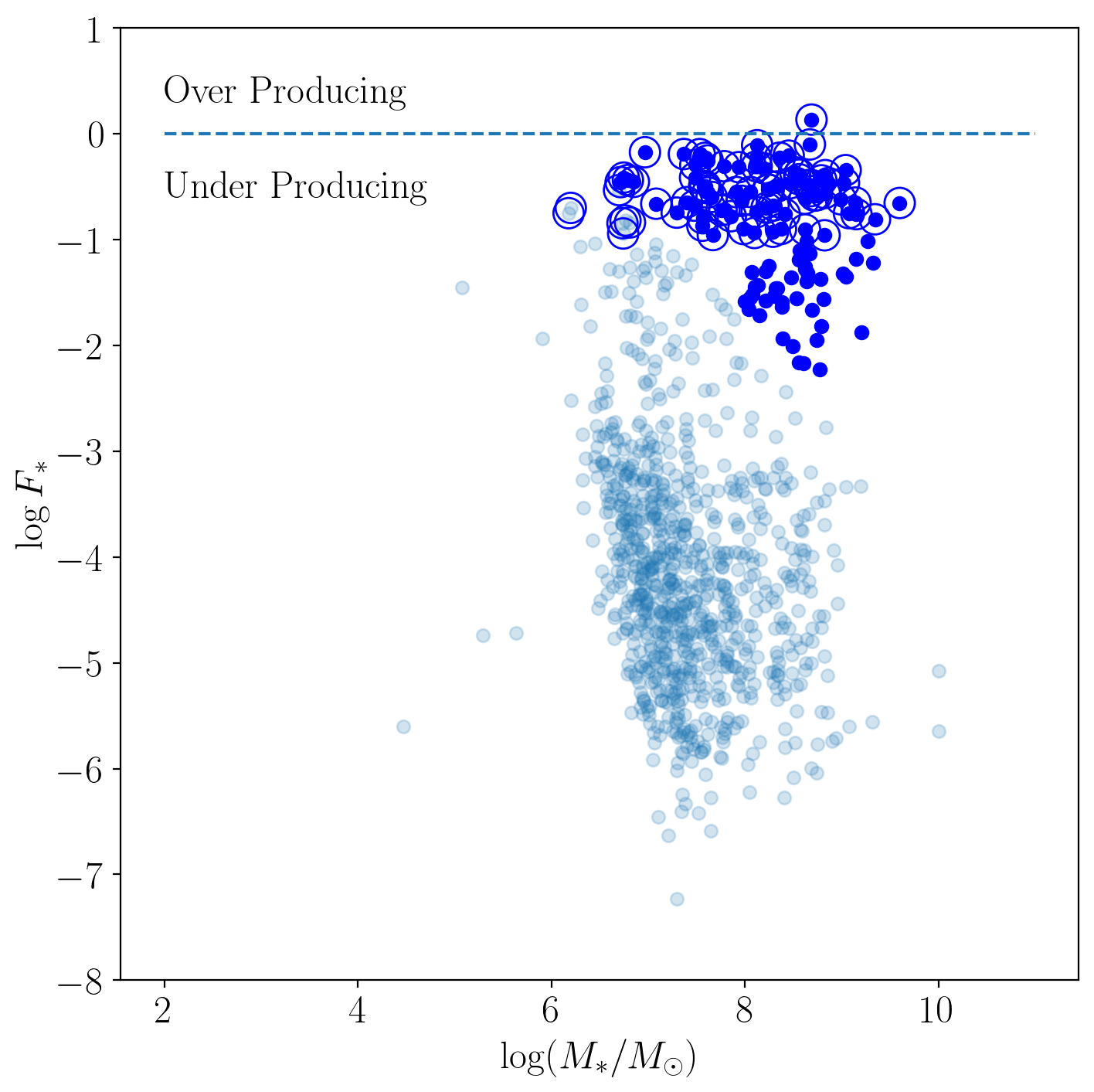}
\end{center}
\vskip -.5cm
\caption{The fraction, $F_*$, of the stellar mass that would have been created if the current SFR had been the SFR over the past  10 Gyr, as a function of stellar mass, $M_*$. 
SMUDGes designated as starforming are highlighted, either as filled blue circles for log(SFR$_0$)$>-3.75$ or larger open circles for log(sSFR$_0$)$>-11$, where sSFR$_0$ is the specific star formation rate (SFR$_0$/$M_*$). The dashed line separates the galaxies for which the current SFR over or underproduces $M_*$.}
\label{fig:frac}
\end{figure}

We examine the nature of SMUDGes with ongoing star formation in Figure \ref{fig:frac}.
Here we present the fraction of the stellar mass, $M_*$, that would have been produced had the galaxy had a constant SFR at the level currently observed for 10 Gyr. We find that in all but one case, the observed level of star formation underproduces the total stars. In other words, the current level of star formation is lower than the average level of star formation even for those SMUDGes that are currently forming stars. For all of these systems, with perhaps the one exception, the dominant epoch of star formation was  in the past. This result is in contrast to what is found for dwIrr galaxies, which appear to have current SFRs that are consistent with being constant over time \citep{hunter}, but consistent with what has been found for H{\small I}-rich UDGs \citep{kado-fong1}.

This is a result that merits closer scrutiny given the results of our simulations. 
Our simulated results suggest 
that the recovered SFR may be biased downward by a few tenths of a dex, depending on the nature of the underlying stellar population. If we add 0.3 dex universally to the measured values of log(SFR$_0$), then
of the 81 systems with log(sSFR$_0) > -11$ (open circles in Figure \ref{fig:frac}), 11, rather than only one, lie above the line. The majority of SMUDGes are still forming stars at a rate below that necessary to create their stars (subcritically efficient), but at least a handful of systems may be critically efficient. Nevertheless, broadly speaking, SMUDGes, whether they are dwarf LSBs or UDGs, had their dominant episode(s) of star formation in the past.

\begin{figure}[ht]
\begin{center}
\includegraphics[width=0.47\textwidth]{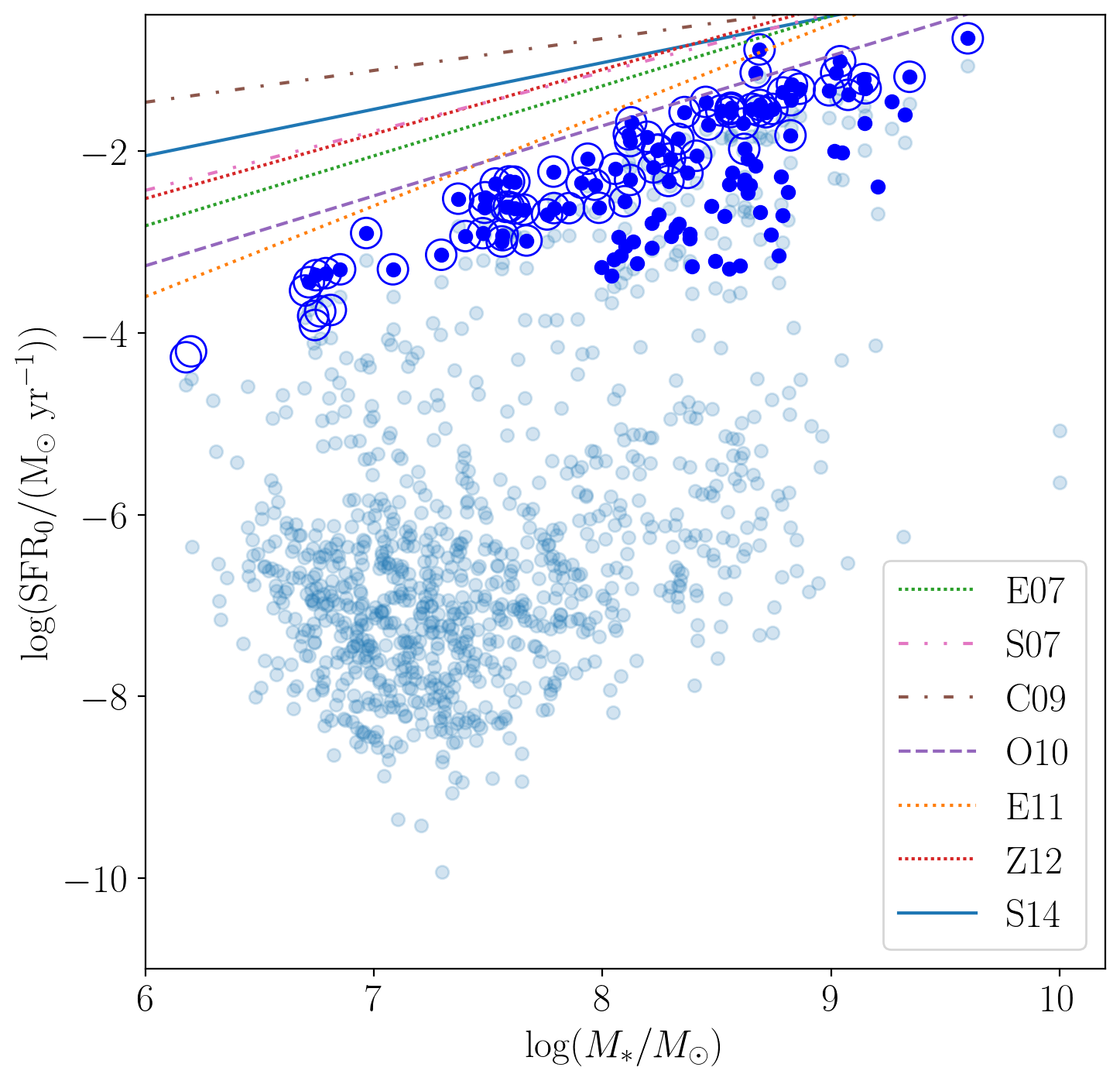}
\end{center}
\vskip -.5cm
\caption{SFR$_0$ vs. $M_*$ for SMUDGes. We compare where the SMUDGes lie relative to various determinations of the galaxy star formation-stellar mass relation. We draw from the compilation of \cite{S14}, who used it to determine the evolution of the relation. We only use the low redshift determinations and adopt a 1 Gyr lookback time for the \cite{S14} formulation. The other relations come from \cite{E07}, \cite{S07}, \cite{C09}, \cite{O10}, \cite{E11} and \cite{Z12}. Galaxies designated as starforming are highlighted, either as filled blue circles for log(SFR$_0$)$>-3.75$ and/or larger open circles for log(sSFR$_0$)$>-11$, where sSFR$_0$ is the specific star formation rate (SFR$_0$/$M_*$). Four galaxies with $\log(M_*/M_\odot) < 6$ are not shown.}
\label{fig:ms}
\end{figure}

Another way of exploring the star forming properties of these galaxies is to place them on the SFR$_0-M_*$ relation. 
This relation is well studied, particularly for more massive galaxies, both locally and at higher redshifts \citep[e.g.,][]{S14}. In Figure \ref{fig:ms} we present SMUDGes in this context and compare to a set of previous determinations of the relation for the star forming sequence of galaxies. All of the published relations are derived from the study of more massive and higher surface brightness galaxies, although with some overlap with the stellar mass range explored here. Despite the evident variation among previous determinations, all of them lie above even the most strongly star-forming SMUDGes (this conclusion is not qualitatively altered by the possible bias discussed above). UDGs are inefficient star-forming systems \citep[see also][]{kado-fong1,kado-fong2}.

\medskip
{\sl Conclusion 2: The star forming systems in SMUDGes are sub-normally efficient. They are not producing stars at a level that puts them on the SFR$_0-M_*$ relation nor at a  rate where they could have produced their current stellar mass over 10 Gyr.}
\medskip

\begin{figure}[ht]
\begin{center}
\includegraphics[width=0.47\textwidth]{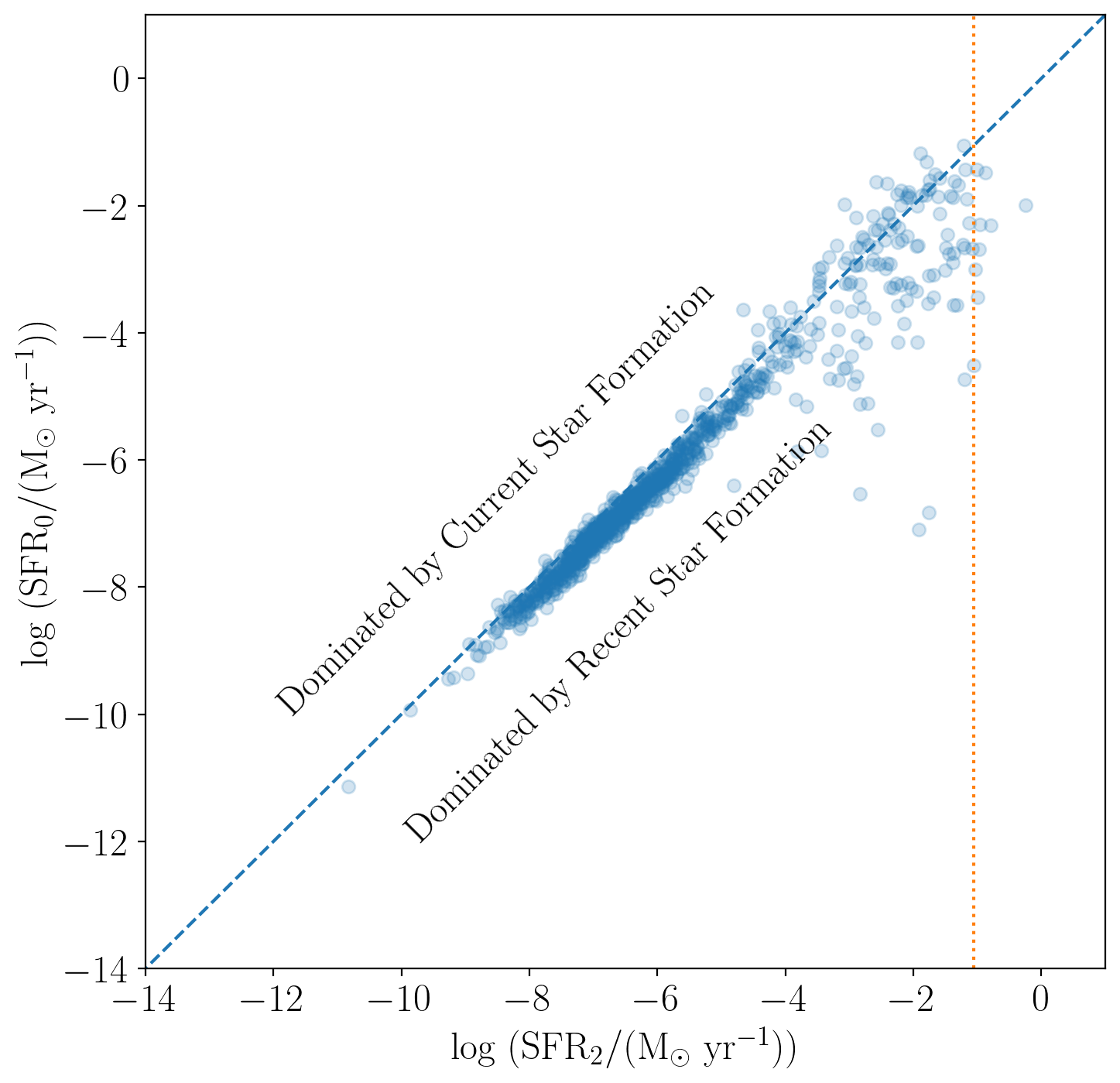}
\end{center}
\vskip -.5cm
\caption{The star formation rates determined for the 0-20 Myr epoch (SFR$_0$) and for the 150 Myr to 1 Gyr epoch (SFR$_2$). The dashed diagonal line is the 1:1 line drawn for guidance and divides the regions of the diagram where galaxies are currently forming stars at a higher mean rate relative to the 150 Myr to 1 Gyr epoch from that where the reverse is the case. To the right of the vertical dotted line the mean star formation rates in the 150 Myr to 1 Gyr epoch were greater than any rate measured currently. Galaxies forming stars at these rates currently are galaxies that are likely to not have met our surface brightness criteria.}
\label{fig:selection}
\end{figure}

SMUDGes are currently low efficiency star forming systems. However, that behavior may not be a universal characteristic of  galaxies in this stellar mass range. Because of our surface brightness selection criteria for SMUDGes ($\mu_{0,g} \ge 24$ mag arcsec$^{-2}$), analogs of SMUDGes within this $M_*$ range that happen to be more strongly star forming may have been excluded from our sample  (see \cite{li} for discussion of this selection bias). We present in Figure \ref{fig:selection} a comparison of the current star formation rate and that inferred for the time interval 150 Myr to 1 Gyr ago, SFR$_2$. If there have been bursts that are significantly stronger in the recent past than what we include in our sample because of our surface brightness selection criteria, we should find some evidence of those in our sample. We do indeed find some systems among the currently star forming systems that had significantly higher (order of magnitude) average SFR between 150 and 1 Gyr ago. These systems are candidates for systems we would have rejected if we had observed them during their dominant star formation episode. Given the age width of the earlier bin, the ``instantaneous" SFR (i.e, over the equivalent 20 Myr window as our first bin) could have been even larger than the average SFRs that we show in Figure \ref{fig:selection}. Given these examples and the assumption that we do not live in a special epoch, it is likely that our sample is  biased in SFR$_0$.

\medskip
{\sl Conclusion 3: SMUDGes galaxies are likely to be biased against strongly star forming systems and not fully representative of the range of SFR$_0$ among galaxies of these stellar masses. Even if UDGs have an intrinsic property that defines their nature, such as high spin halos, these results suggest that some may not currently satisfy the UDG criteria.}
\medskip

The possibility of stronger star formation episodes in the past raises the question of whether unquenched UDGs are distinct from other field galaxies of similar total mass or are simply being caught at a particular time during a star formation cycle. We cannot answer this with the data presented here, but previous studies \citep{Zaritsky_2023,forbes25} concluded that the stellar-to-halo mass ratios of UDGs in general are lower than average for galaxies of their halo mass. As such, SMUDGes that are UDGs, not only have low current SFRs but also have lower than typical average SFR.

\medskip
{\sl Conclusion 4: The SMUDGes galaxies include some that have had significantly higher SFRs in the recent ($<$ 1 Gyr) past but have faded sufficiently in surface brightness that they currently satisfy the UDG surface brightness criterion. As such, we do find significant variations in the SFR on timescales $< 1$ Gyr. Nevertheless, we do not support a scenario in which the UDG criterion is simply catching normal galaxies in a low-SF phase.}

\medskip

\begin{figure}[ht]
\begin{center}
\includegraphics[width=0.47\textwidth]{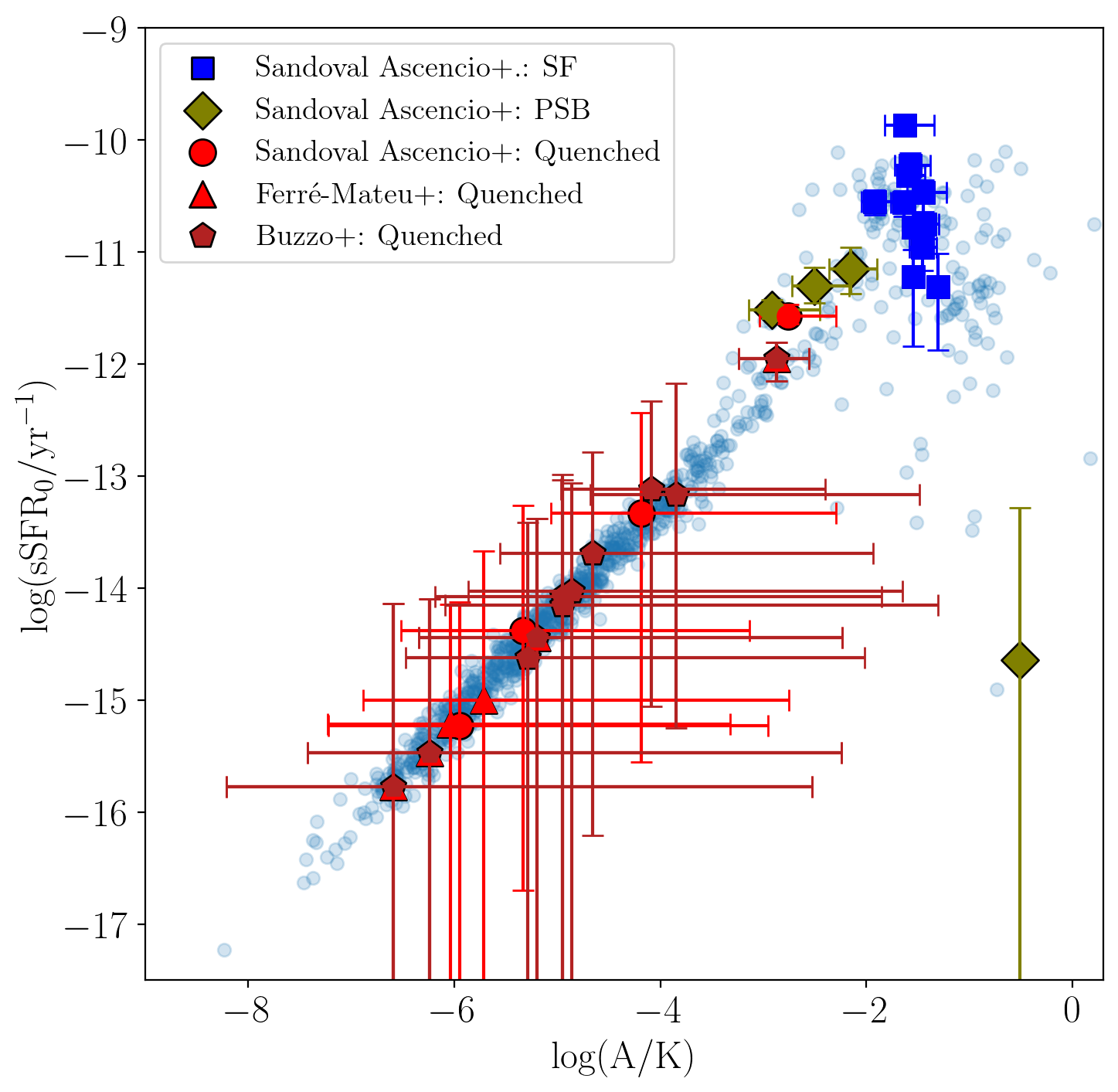}
\end{center}
\vskip -.5cm
\caption{A/K plotted against specific star formation rate, sSFR$_0$, as measured from our PROSPECTOR modeling for SMUDGes galaxies. The larger, filled points highlight the same objects as in Figure \ref{fig:evol}.}
\label{fig:ssfr}
\end{figure}

At each value of A/K in Figure \ref{fig:evol} there is significant scatter in SFR$_0$. One reason for this scatter, a dependence on stellar mass, was already proposed when we compared LSBs and UDGs within the sample. The straightforward way to remove this behavior is by calculating the specific SFR$_0$ (sSFR$_0$), which is simply SFR$_0$ normalized by the galaxy's stellar mass, $M_*$. This approach also removes the effect of distance errors as a potential source of scatter. We present the results for sSFR$_0$ vs. A/K in Figure \ref{fig:ssfr}. 

The relationship tightens significantly, confirming that the source of most of the original scatter was due to the wide range of stellar masses in the sample. The relationship shows a sharp increase in sSFR$_0$ as A/K increases, as one might expect in systems that continue to form stars for a time that is long enough to link the current star formation rate with the number of intermediate-mass stars, $\sim 1$ Gyr. However, there is a somewhat surprising reversal in that trend beyond the peak at log(A/K) $\sim -2$ toward the largest values of A/K. We will return to this feature later. Among the quenched galaxies, there also appears to be an exceedingly tight relationship between A/K and sSFR, but the large error bars argue against assigning physical significance to this relation. Instead, for quenched galaxies we suspect that errors in M$_*$, which is dominated by $M_3$, lead to galaxies sliding along the diagonal because both axes are affected.
Given the uncertainties, the quenched galaxies are mostly statistically indistinguishable from each other in terms of sSFR$_0$ or A/K.

\medskip
{\sl Conclusion 5: The SMUDGes galaxies exhibit more coherent behavior when the specific star formation rate is examined. There is a reversal in the general behavior between sSFR$_0$ and A/K for large values of A/K that merits further investigation and a tight behavior in these properties for quenched galaxies that is not physical.}
\medskip

\subsection{Recent Star Formation}
\label{sec:recentSF}

Our Conclusion 4 suggests that the recent SFH of these galaxies might show significant fluctuations. Again, we need to turn to simulations to better understand what the PROSPECTOR analysis is able to faithfully recover in such cases. We show in Appendix \ref{app:sims} that we have varying degrees of success in identifying different possible variations. We will place those results in context as we proceed in this discussion.

We focus here on our previously defined A/K index.
One striking aspect of Figure \ref{fig:evol} is the absence of galaxies with high SFR$_0$ and low A/K values. This apparent result suggests that we have no long-term quenched galaxies that have recently and suddenly re-activated star formation. Such a finding would have implications for how long a galaxy could retain the fuel for star formation without forming stars or for how a galaxy might re-acquire that fuel. As such, the absence (or presence) of such objects is of scientific interest.

The interpretation of this result is predicated on the ability of the SFH reconstruction to recover a current burst without ``leakage" to intermediate age bins (otherwise the galaxy would appear to have higher A/K and merge in the Figure with other star forming galaxies). 
Our simulations (Appendix \ref{app:sims}) show that our analysis methodology can miss such objects if they exist. As such, the absence of such objects in Figure \ref{fig:evol} cannot be used to interpret that they are truly absent. This is unfortunate, but it is a strict limitation of our approach.

\medskip
{\sl Conclusion 6: Galaxies that have been otherwise quiescent for longer than 1 Gyr but have a current sudden increase in star formation can be misidentified in our analysis. Simulations are critical in determining which features in this parameter space are recoverable and which are not.}

\medskip
We now focus on the complementary question of whether we can identify a galaxy that has had significant recent star formation but has little or no current star formation (i.e., a post-starburst). We have already discussed that if a galaxy is currently forming stars, we can recover the current star formation rate well.
We find (Appendix \ref{app:sims}) that galaxies that have a large fraction of intermediate age stars, A/K $\sim 1$, and low to negligible current star formation rates are correctly identified as such systems. However, we tend to overestimate the current star formation rates, particularly for those with exceedingly low input star formation rates. As such, we cannot expect to identify systems that are fully quenched at the current time as such (they will appear to have some residual current star formation). To identify PSBs we should set a modest upper limit on the star formation rate rather than require the apparent complete cessation of star formation.

\medskip
{\sl Conclusion 7: Our methodology  can identify galaxies that have had a recent (age $< 1$ Gyr) episode of star formation and are currently forming few if any stars (log(sSFR$_0$/yr$^{-1}) < -11$). Our sample includes such systems.}

\subsection{Classifying Galaxies}

\begin{figure}[ht]
\begin{center}
\includegraphics[width=0.47\textwidth]{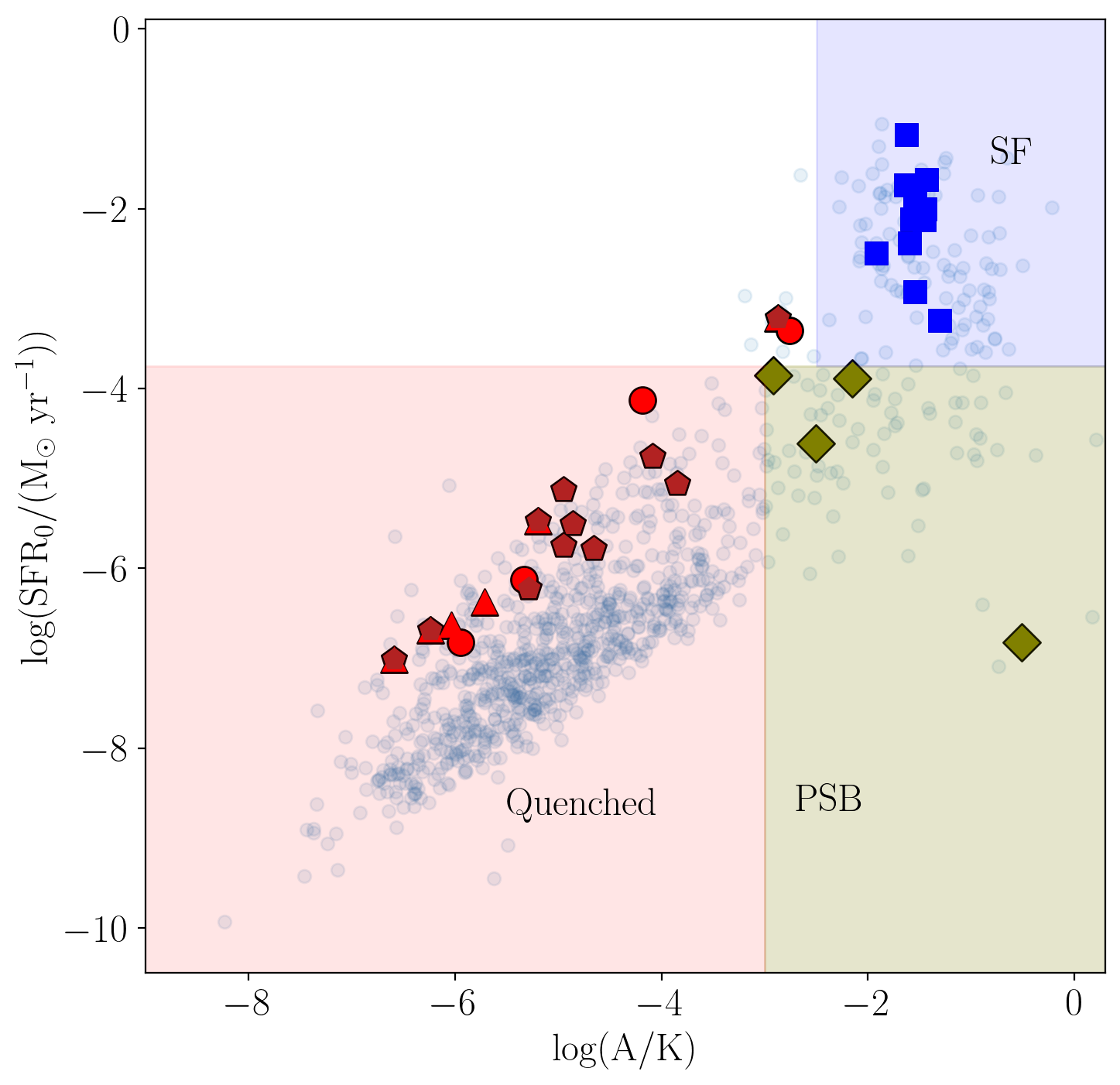}
\end{center}
\vskip -.5cm
\caption{Categorization regions. A reprise of Figure \ref{fig:evol} with regions highlighted where we define star forming (SF), post-starburst (PSB), and quenched galaxies. Larger symbols again highlight the independently classified galaxies that are also highlighted in Figures \ref{fig:evol} and \ref{fig:ssfr}. We have omitted the errorbars for clarity but they are the same as in Figure \ref{fig:evol}.}
\label{fig:categorize}
\end{figure}

We categorize the galaxies, with guidance from the spectroscopic classifications of \cite{Ferre-Mateu+2018} and \cite{loraine}, and from the SED results of \cite{Buzzo_2022} in Figure \ref{fig:categorize}.
We identify post-starburst, or at least post-starforming, SMUDGes by requiring both a high value of A/K ($-3 <$ log(A/K)) and a low SFR$_0$ (log(SFR$_0$) $< -3.75$). There are 74 candidate PSB galaxies using these criteria in this sample.
For quenched galaxies we require (log(A/K) $< -3$) and a low SFR$_0$ (log(SFR$_0$) $< -3.75$), while for star forming galaxies
($-2.5 <$ log(A/K)) and a high SFR$_0$ ($-3.75 <$ log(SFR$_0$) ).

\begin{figure}[ht]
\begin{center}
\includegraphics[width=0.47\textwidth]{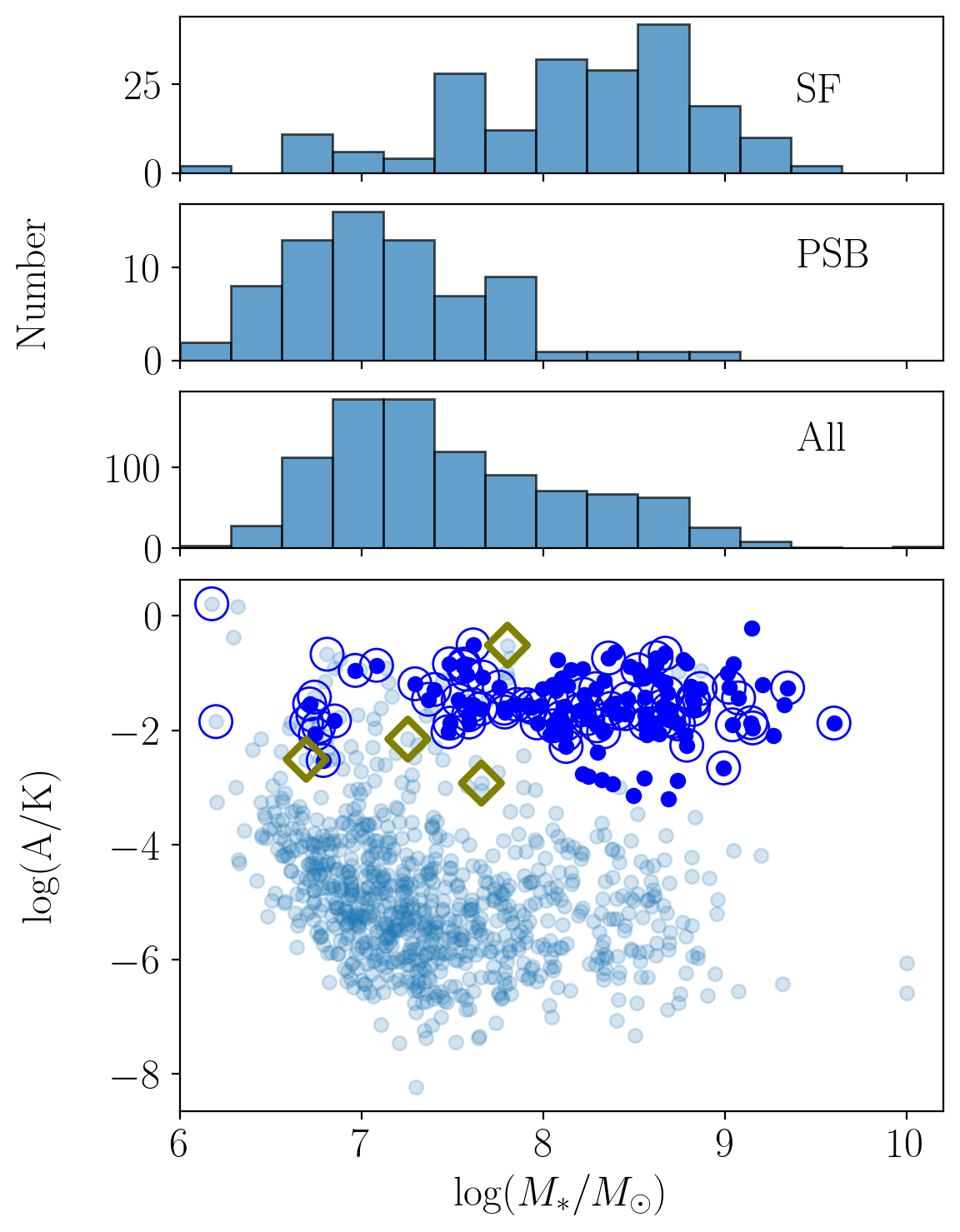}
\end{center}
\vskip -.5cm
\caption{The intermediate age diagnostic, A/K, as a function of stellar mass, $M_*$. Bottom panel shows the distribution with three types of objects highlighted: 1) those with log(sSFR$_0) > -11$ (filled blue circles), 2) those with log(SFR$_0$) $>-3.75$ (larger unfilled circles), and 3) those identified by \cite{loraine} as post-starburst (PSB; filled green diamonds). In the upper panel we show the distribution in $M_*$ for the star forming systems (those with either filled blue or open blue circles). In the second panel down we show the distribution for those systems that satisfy  log(SFR$_0$) $<-3.75$ and log(A/K) $> -3$, which is our PSB criteria. In the third panel down we show the distribution for the entire sample. PSB candidates may have a slight bias toward less massive systems but there are some candidates even at the largest masses we probe. The PSB and SF distributions differ significantly.}
\label{fig:ak}
\end{figure}

Using these selection criteria, we compare the stellar mass distribution of the SF and PSB galaxies to that of the full sample in Figure \ref{fig:ak}. We find candidate PSBs across the full range of stellar mass in our sample but with an apparent preference for the lower masses. Nevertheless, we caution against drawing strong inferences from the distributions of subclasses because of the various biases in selecting this sample. It is evident from Figure \ref{fig:ssfr} that selection criteria based on sSFR$_0$ would not produce a clean sample of PSBs. What our selection criteria do provide is a sample of objects in the various classes that can be studied further, rather than a complete set of each class.

\medskip
{\sl Conclusion 8: PSB galaxies are found in our sample at all stellar masses, but there is a preference, given our selection criteria, for them to be among the lower mass systems. In contrast, star forming systems lie preferentially toward the upper mass range of our sample. There are potential biases in the sample that preclude a definitive interpretation but the result is consistent with the supposition that higher mass galaxies can maintain a more regular rate of SF while lower ones are more cyclical. This is in line with results seen generally among galaxies \citep{searle}.}
\medskip

Surrounding various of these questions is the unknown nature of star forming galaxies that might lie above the SMUDGes surface brightness limit. As we have noted previously, it is likely that there is a population of star forming objects with the same range of stellar masses that lies just above our surface brightness cut and are otherwise comparable to SMUDGes. This bias highlights the challenge of understanding the full population of systems when a strict parameter cut is applied. In such cases, it is advisable to forward model the sample using the selection that was imposed, rather than analyzing the existing sample to draw conclusions. 

\subsection{Do Star Formation Episodes Lead to Larger Galaxies?}

\begin{figure}[ht]
\begin{center}
\includegraphics[width=0.47\textwidth]{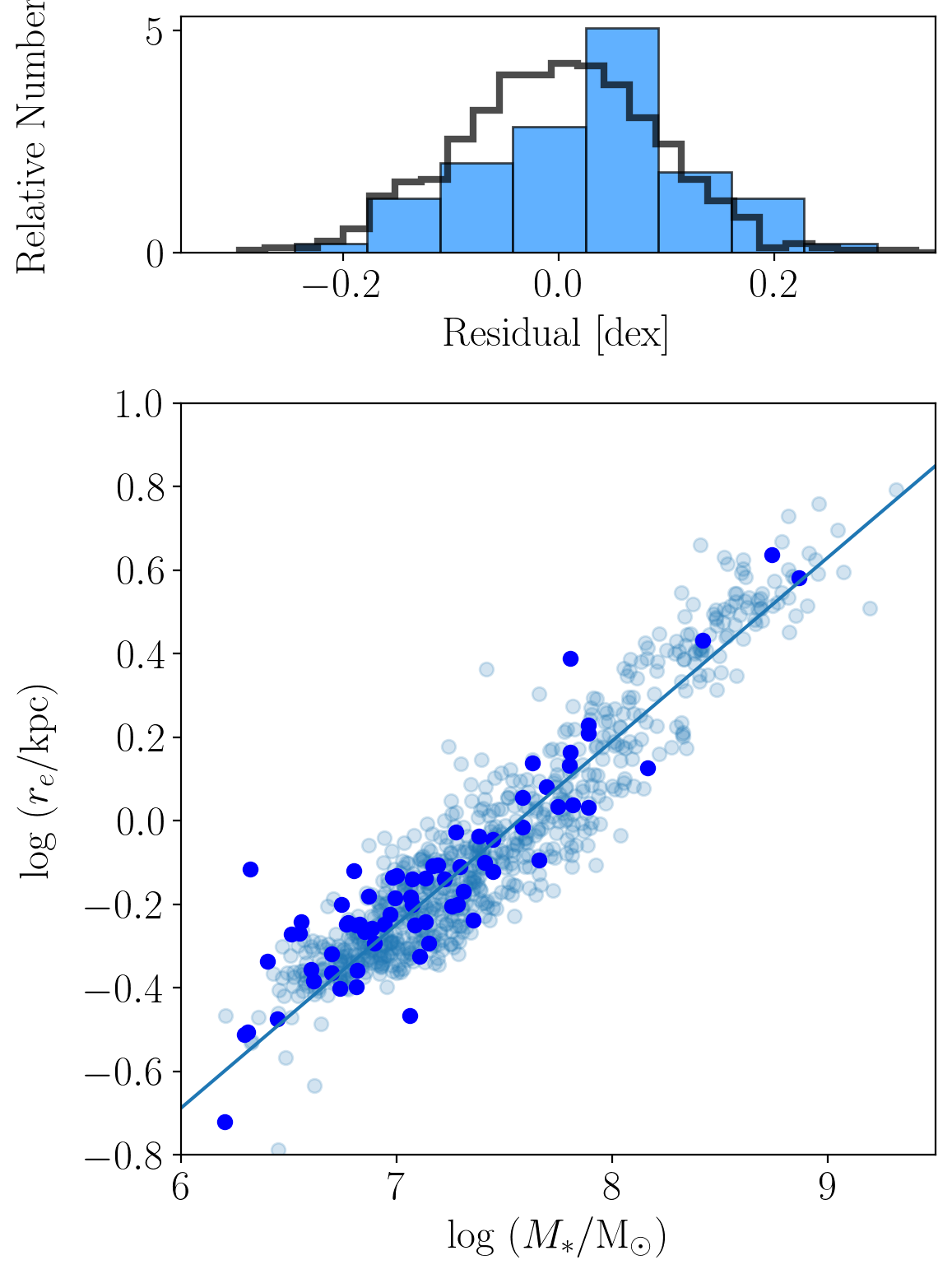}
\end{center}
\vskip -.5cm
\caption{The scale length - stellar mass relation for the quenched SMUDGes samples and for the subsample of PSB galaxies (filled circles). There is a minor shift to larger sizes in the PSB galaxies at a given stellar mass ($0.034 \pm 0.012$ dex, or $\sim$ 8\%) seen in the upper panel where we plot residuals relative to the best fit relationship from the lower panel. The histograms in the upper panel are normalized relative to each other (filled histogram represents the PSB sample, solid line the quenched SUMDGes sample).} 
\label{fig:re}
\end{figure}

Next we probe whether there is any indication that a star formation episode leads to an enlargement of the galaxy. In Figure \ref{fig:re} we examine whether those galaxies with a recent episode, those we classify as PSB, are on average larger than their quenched counterparts. To do this, we normalize using the $r_e$-$M_*$ relation and examine the distribution of the residuals about this relationship for the sample in general and for PSBs. 
We adopt and fit a power law relationship between $r_e$ and $M_*$, finding $r_e =({M_*/(3.32\times 10^7})^{0.44}$, where $r_e$ is in kpc and $M_*$ in solar units.

We find that the PSBs are slightly larger, offset from the mean relation by 
$0.034 \pm 0.012$ dex (or 8\% larger). The probability of such a difference in the mean occurring randomly is 0.004, as determined using a standard t-test. Of course, this shift could be both statistically significant and physically insignificant. For example, it could be exaggerated due to a slight variation in the locations where the younger stars formed relative to the older stars if the growth of these galaxies is inside-out. Our estimate of $r_e$ would be skewed large by the higher luminosity of these younger stars, but would ease back to the previous value as this younger stellar population fades.  Higher angular resolution IR imaging, with which sizes could be measured while minimizing the effect of the younger stellar population, is needed to address this concern. In more massive PSBs, the trend for where the new stars form is in the opposite sense \citep{norton}, but those systems appear to be mostly driven by interactions \citep{zabludoff} rather than the cyclical pattern of star formation that we suspect to be at work here. On the other hand, if this measurement represents galaxy growth by 8\%, then nine such episodes would double the size of the galaxy. If these episodes happen every $\sim$ Gyr, there is sufficient time for an effect of this magnitude. As we have also noted, at least some of the previous star formation episodes had to be of greater magnitude than the current episodes we are witnessing, so fewer episodes might be needed to have a comparable effect. This may not be a viable model for the formation of cluster UDGs if their star formation was quenched by early infall into the cluster, but it might be viable for field UDGs if one can explain why such episodes occur only in some dwarf galaxies and address the issues raised \cite{kado-fong1}. They argued on both empirical grounds, based on the current extended distribution of star formation, and theoretical grounds, a disagreement with predictions from bursty galaxy models, that feedback driven expansion does not appear to be the responsible mechanism for the character of UDGs, unless that feedback happened at high redshift.

\medskip
{\sl Conclusion 9. There is a suggestion in the data that star formation episodes marginally increase (by $\sim 8\%$) the effective radius of galaxies in our sample. This is likely to be an upper limit, or perhaps just an observational bias driven by where the recent star formation occurs. Taken at face value, the result suggests that order one changes to the size require either multiple episodes or much more violent ones than the ones we are observing in these galaxies.}

\medskip
Of course, there is always a danger of oversimplification. Various mechanisms are possibly at work creating what is a heterogeneous population of UDGs. For example, \cite{dicintio+2019} suggest that feedback dominates for lower mass UDGs and high angular momentum for higher mass UDGs. In their models the transition in behavior occurs at galaxy stellar masses of $\sim 10^9$ M$_\odot$, which is above the stellar mass of almost all of our SMUDGes, but our cautionary note still holds.

\medskip

\section{Summary}

We present UV and IR photometry for the SMUDGes catalog, which we then use to construct the spectral energy distribution (SED) and recover the recent star formation history (SFH) for the 966 sources from the catalog for which we have either a spectroscopic or estimated redshift. We present both the photometry and the recovered star formation rates (SFR) as a function of age in tabular form. We find no clear differences between galaxies with spectroscopic or estimated redshifts or large (UDG) or small (normal dwarf) galaxies and so we present and analyze the full sample.

We find that our SED analysis is able to identify star forming, post-starburst, and quenched galaxies at a precision comparable to what has been done with IFU spectroscopy. This enables us to identify 74 candidate LSB PSB galaxies, including 6 that satisfy the UDG size criterion\footnote{Different studies adopt different central surface brightness and/or size criteria; for example, \cite{loraine} adopted $r_e > 1 $ kpc as their UDG size criterion. For that particular choice, we find 15 PSB UDGs in our sample.}. 

We find that the star forming systems among our LSB population are less efficient at forming stars over the last 20 Myr than are normal star forming galaxies. They are neither producing stars at a level that places them on the SFR-$M_*$ relation nor at a SFR where they could have produced their current stellar mass over 10 Gyr if the rate remained constant. We conclude that they must have had more vigorous star formation episodes in the past. However, we caution that some more vigorous star forming galaxies of similar stellar mass would have most likely been excluded from our sample by the SMUDGes surface brightness selection criterion. As such, weak star formation is a property of the SMUDGes sample, not necessarily of galaxies of this stellar mass or size.

We find that our analysis produces false and true features in the diagnostic diagrams of SFR$_0$ or sSFR$_0$ vs. A/K. Simple simulations that we carry out to test our methodology show us that the absence of quenched galaxies in our sample that have recently re-activated star formation (log(A/K) $\lesssim -4$ and log(sSFR$_0$) $> -12.5$) may be a mirage. Our methodology tends to place such objects, when they exist, at different locations in the diagram. However, the downturn in the relationship between sSFR$_0$ and A/K for 
log(A/K) $> -2$ is true. We find that the analysis robustly returns sSFR$_0$ and can return the properties of PSB galaxies reliably.

The identification of PSB galaxies demonstrates that SFRs vary significantly over the last Gyr in a number of SMUDGes systems. We cannot determine whether those systems with recent star formation that are currently quenched will remain so, or if they are part of a cyclical process where SF will begin anew. However, the number of such systems suggests that this is unlikely to be their last episode.

We find
PSB galaxies in our sample at all stellar masses, but with a preference among the lower mass systems. In contrast, star forming systems lie preferentially toward the upper mass range of our sample. There are potential biases caused by our selection criteria that could affect these distributions and preclude a definitive interpretation. Nevertheless, the result is consistent with the supposition that higher mass galaxies can maintain a more regular rate of SF while lower ones are more cyclical.

We find evidence that star formation episodes marginally increase (by $\sim 8\%$) the size of galaxies in our sample. If this is not merely an observational bias caused by star formation occurring at larger radii, this is still  likely to be an upper limit to the size change per episode and as such suggests that order one changes to the size require either multiple episodes or much more violent ones.

The availability of the recent star formation histories for a large sample of low surface brightness galaxies enables  investigations of a number of additional topics beyond those explored here. We are in the process of examining connections between environment and recent SFH (Khim et al. in prep.), the presence of a nuclear cluster \citep{lambert,khim} and recent SFH, and H{\small I} properties \citep{karunakaran24,motiwala} and recent SFH (Sandoval Ascencio et al. in prep.) By making both the photometry and our modeled SFHs available, we aim to enable others to explore these and other questions as well.

\begin{acknowledgments}

We thank an anonymous referee for helpful comments that improved the manuscript.
DZ, RD, and DJK acknowledge financial support from NSF AST-1713841, AST-2006785, and AST-2510821 and NASA 22-ADAP-011. KS acknowledges funding from the Natural Sciences and Engineering Research Council of Canada (NSERC). An allocation of computer time from the University of Arizona Research Computing High Performance Computing (HPC) Research Computing group and the prompt assistance of the associated computer support group are gratefully acknowledged.

This research has used the NASA IPAC Extragalactic Database (NED), which is operated by the Jet Propulsion Laboratory, California Institute of Technology, under contract with NASA.

The Legacy Surveys consist of three individual and complementary projects: the Dark Energy Camera Legacy Survey (DECaLS; Proposal ID \#2014B-0404; PIs: David Schlegel and Arjun Dey), the Beijing-Arizona Sky Survey (BASS; NOAO Prop. ID \#2015A-0801; PIs: Zhou Xu and Xiaohui Fan), and the Mayall z-band Legacy Survey (MzLS; Prop. ID \#2016A-0453; PI: Arjun Dey). DECaLS, BASS and MzLS together include data obtained, respectively, at the Blanco telescope, Cerro Tololo Inter-American Observatory, NSF's NOIRLab; the Bok telescope, Steward Observatory, University of Arizona; and the Mayall telescope, Kitt Peak National Observatory, NOIRLab. The pipeline processing and analysis of the data were supported by NOIRLab and the Lawrence Berkeley National Laboratory (LBNL). The Legacy Surveys project is honored to be permitted to conduct astronomical research on Iolkam Du'ag (Kitt Peak), a mountain with particular significance to the Tohono O'odham Nation.

NOIRLab is operated by the Association of Universities for Research in Astronomy (AURA) under a cooperative agreement with the National Science Foundation. LBNL is managed by the Regents of the University of California under contract to the U.S. Department of Energy.

This project used data obtained with the Dark Energy Camera (DECam), which was constructed by the Dark Energy Survey (DES) collaboration. Funding for the DES Projects has been provided by the U.S. Department of Energy, the U.S. National Science Foundation, the Ministry of Science and Education of Spain, the Science and Technology Facilities Council of the United Kingdom, the Higher Education Funding Council for England, the National Center for Supercomputing Applications at the University of Illinois at Urbana-Champaign, the Kavli Institute of Cosmological Physics at the University of Chicago, Center for Cosmology and Astro-Particle Physics at the Ohio State University, the Mitchell Institute for Fundamental Physics and Astronomy at Texas A\&M University, Financiadora de Estudos e Projetos, Fundacao Carlos Chagas Filho de Amparo, Financiadora de Estudos e Projetos, Fundacao Carlos Chagas Filho de Amparo a Pesquisa do Estado do Rio de Janeiro, Conselho Nacional de Desenvolvimento Cientifico e Tecnologico and the Ministerio da Ciencia, Tecnologia e Inovacao, the Deutsche Forschungsgemeinschaft and the Collaborating Institutions in the Dark Energy Survey. The Collaborating Institutions are Argonne National Laboratory, the University of California at Santa Cruz, the University of Cambridge, Centro de Investigaciones Energeticas, Medioambientales y Tecnologicas-Madrid, the University of Chicago, University College London, the DES-Brazil Consortium, the University of Edinburgh, the Eidgenossische Technische Hochschule (ETH) Zurich, Fermi National Accelerator Laboratory, the University of Illinois at Urbana-Champaign, the Institut de Ciencies de l'Espai (IEEC/CSIC), the Institut de Fisica d’Altes Energies, Lawrence Berkeley National Laboratory, the Ludwig Maximilians Universitat Munchen and the associated Excellence Cluster Universe, the University of Michigan, NSF’s NOIRLab, the University of Nottingham, the Ohio State University, the University of Pennsylvania, the University of Portsmouth, SLAC National Accelerator Laboratory, Stanford University, the University of Sussex, and Texas A\&M University.

BASS is a key project of the Telescope Access Program (TAP), which has been funded by the National Astronomical Observatories of China, the Chinese Academy of Sciences (the Strategic Priority Research Program “The Emergence of Cosmological Structures” Grant \#XDB09000000), and the Special Fund for Astronomy from the Ministry of Finance. The BASS is also supported by the External Cooperation Program of Chinese Academy of Sciences (Grant \#4A11KYSB20160057), and Chinese National Natural Science Foundation (Grant \#12120101003, \#11433005).

The Legacy Survey team makes use of data products from the Near-Earth Object Wide-field Infrared Survey Explorer (NEOWISE), which is a project of the Jet Propulsion Laboratory/California Institute of Technology. NEOWISE is funded by the National Aeronautics and Space Administration.

The Legacy Surveys imaging of the DESI footprint is supported by the Director, Office of Science, Office of High Energy Physics of the U.S. Department of Energy under Contract No. DE-AC02-05CH1123, by the National Energy Research Scientific Computing Center, a DOE Office of Science User Facility under the same contract; and by the U.S. National Science Foundation, Division of Astronomical Sciences under Contract No. AST-0950945 to NOAO.

This research used data based on observations made with the Galaxy Evolution Explorer, obtained from the MAST data archive at the Space Telescope Science Institute, which is operated by the Association of Universities for Research in Astronomy, Inc., under NASA contract NAS 5–26555. The specific observations analyzed can be accessed via \dataset[https://doi.org/10.17909/npwd-sh81]{https://doi.org/10.17909/npwd-sh81}. 

\end{acknowledgments}

\medskip
\facilities{GALEX, WISE}

\software{
Astropy              \citep{astropy:2013, astropy:2018, astropy:2022},
astroquery           \citep{astroquery},
asymmetric$\_$uncertainty \citep{asymuncert},
dustmaps
\citep{green},
dynesty
\citep{speagle,koposov},
fsps
\citep{conroy2009, conroy2010},
GALFIT               \citep{peng},
h5py
\citep{h5py},
Matplotlib           \citep{matplotlib},
NumPy                \citep{numpy},
pandas               \citep{pandas},
PROSPECTOR           \citep{prospector},
SciPy                \citep{scipy1, scipy2},
sep                  \citep{sep}}

\bibliography{refs.bib}
\bibliographystyle{aasjournal}

\begin{appendix}

\section{Comparing Our Photometry to Published Values}

Several smaller prior studies have investigated the SFHs of UDGs.  In most cases, their methodology differs significantly from ours in terms of both their photometry and  modeling. Here we compare to their photometry to provide an external check on the robustness of the measurements.

\subsection{UV Photometry}

\begin{figure*}[ht]
\includegraphics[width=1.0\textwidth]{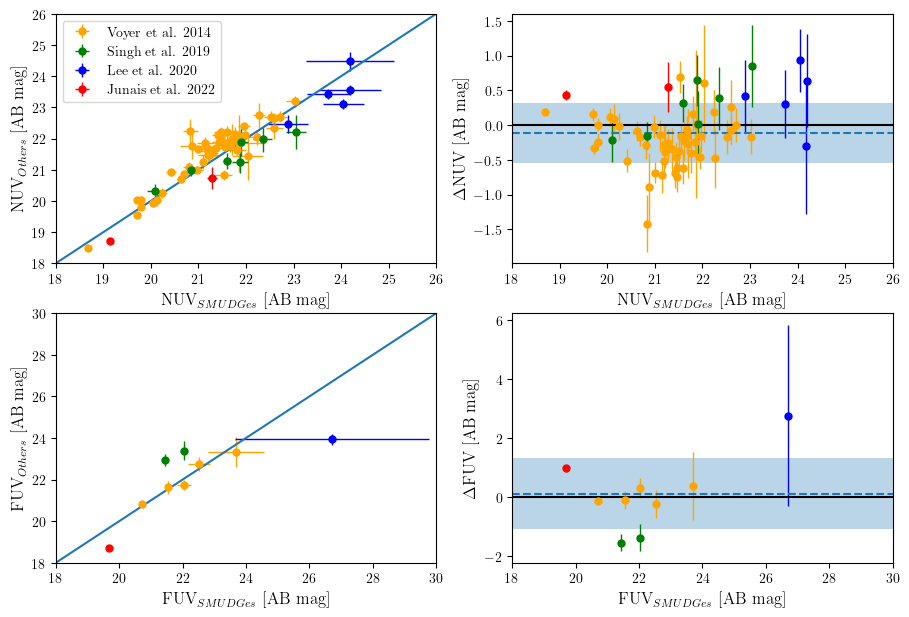}
\caption{Comparisons between UV mags estimated by other authors and those obtained in this study.  The left panels show the direct comparison while the right ones show the differences (SMUDGes - Others).  Dashed lines in the right panels indicate the mean differences of $-0.11\pm0.05$ for NUV and $0.12\pm0.40$ mag for FUV. Individual measurement standard deviations are 0.43 and 1.20 mag, respectively, and shown as the shaded regions.}
\label{fig:GALEXphot}
\end{figure*}

Several authors have measured UV fluxes of UDGs \citep{voyer_2014,Singh_2019,Lee+2020,Junais_2022}. Generally, the approach has been similar to ours, using aperture photometry but differing in how the apertures are defined.  We present comparisons for galaxies in common among these studies in Figure \ref{fig:GALEXphot}. The apparent magnitudes agree  well, with small mean zero point offsets (NUV, FUV = $-0.11\pm0.05$, $0.12\pm0.40$ mag). 

We now describe the individual literature studies and any key differences.
\cite{voyer_2014} limited their UV estimates to sources in the Virgo cluster observed by GALEX. Although they used apertures that encompassed the entire galaxy, 
the overall agreement is good with no evidence of significant systematic biases.
\cite{Singh_2019} used an earlier version of the SMUDGes catalog (Paper I) using a circular aperture with a radius of 2$r_e$ estimated with a fixed S\'ersic index of one. Their estimates of $r_e$ differ from our more recent values and, together with the use of circular apertures and different methods to estimate background contamination, probably account for the magnitude offset.  
As shown in Figure \ref{fig:GALEXphot}, many of our UDGs that match those of \cite{Lee+2020} are faint and at the edge of GALEX detectability, so, as expected, uncertainties in UV photometry are relatively large.  They also used a circular aperture, but used a radius of 1$r_e$ obtained from \cite{vanDokkum+2015}. For the object with the largest offset in NUV (DF25,  SMDG1259487+274639), they adopted $r_e=9.4$\arcsec compared to a SMUDGes value of 7.0\arcsec.  Despite these differences, the magnitude difference is not highly significant. 
\cite{Junais_2022} estimated UV flux in UDGs located within the Virgo cluster. They generated their apertures from the largest measured radius in their g-band images, which is 3$\sigma$ above the sky level.  Our results have an average offset of $\sim$0.5 mag in NUV, although there are only two galaxies for comparison. We conclude that we find no systematic differences between our UV photometry and the averages of those in the literature that are greater than our internal error estimates. 

\subsection{IR Photometry}
\label{app:ir}
\begin{figure*}[ht]
\includegraphics[width=1.0\textwidth]{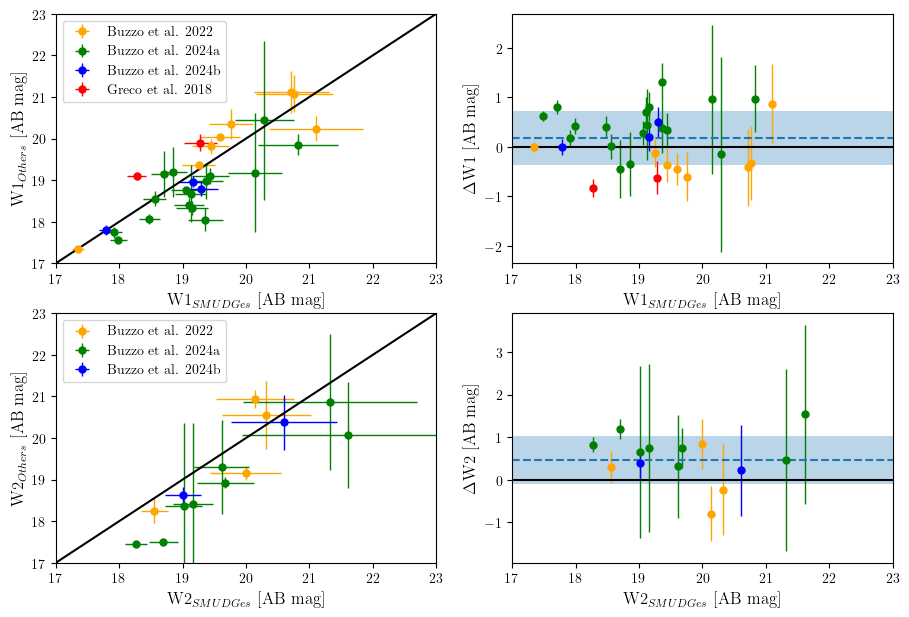}
\caption{Comparisons between WISE mags estimated by other authors and those obtained in this study.  Plots are as described in Figure \ref{fig:GALEXphot}. We find mean differences of $0.18\pm0.10$ mag for W1 and $0.47\pm0.14$ mag for W2.
Individual measurement standard deviations are 0.54 and 0.56 mag, respectively, and shown as the shaded regions.}
\label{fig:WISEphot}
\end{figure*}

\begin{figure*}[ht]
\includegraphics[width=1.0\textwidth]{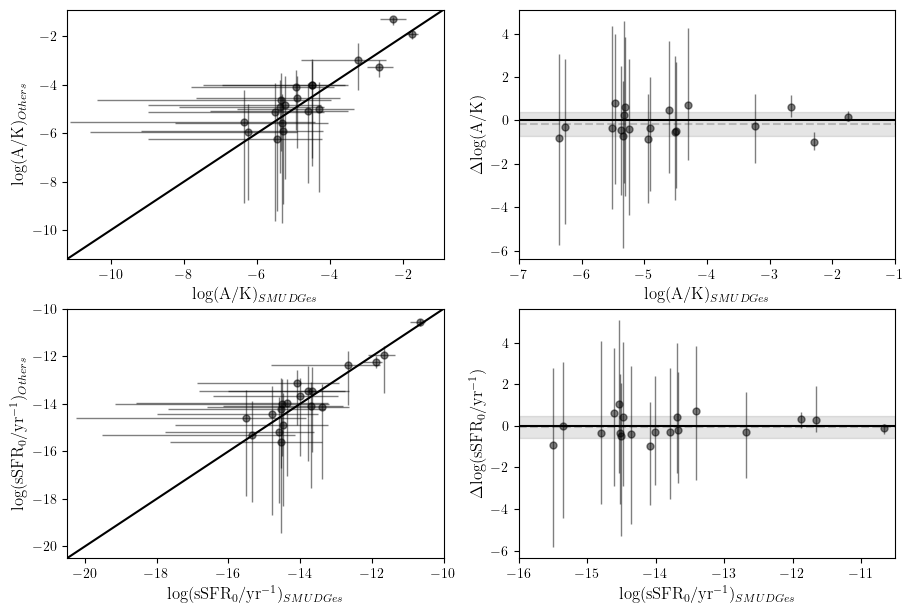}
\caption{Changes in log(A/K) and log(sSFR$_0$) after replacing original W1 and W2 photometry with values obtained from other sources. Dashed lines in the right panels indicate the mean differences (SMUDGes $-$ Others) which are $-0.16\pm0.13$ for log(A/K) and $-0.05\pm0.12$ for log(sSFR$_0$). Standard deviations for individual measurements, shown as the shaded regions, are 0.56 and 0.53, respectively. Uncertainties for the differences are generated using the Python package asymmetric$\_$uncertainty \citep{asymuncert}.}
\label{fig:WISEprospect}
\end{figure*}

Other studies examining the SFHs of UDGs have also taken advantage of WISE IR imaging \citep{Buzzo_2022, buzzo2024a, buzzo2024b, Greco+2018b}.  However, in all cases their methodologies differed significantly from ours in both how they acquired the images and the photometry obtained. For example, \cite{Buzzo_2022, buzzo2024a, buzzo2024b} obtained their images from archival ALLWISE data as well as mosaics created from single frames, while, as noted in Section \ref{subsubsec:photIR}, we downloaded ours as coadded bricks from the Legacy Survey. These discrepancies are further compounded by the relatively poor resolution of WISE images \citep[6.1, 6.4, 6.5, and 12.0\arcsec\ for W1-4, respectively,][]{Wright_2010}, the faintness of these particular galaxies, and the crowded fields in which they frequently lie. Comparing the WISE results for W1 and W2 (Figure \ref{fig:WISEphot}), we find that even with significant processing differences, our results correlate well (r = 0.87 for W1 and 0.86 for W2) and have mean zero point magnitude offsets of $0.18\pm0.10$ and $0.47\pm0.14$ mag for W1 and W2, respectively, relative to the complete set of data from the literature.  There were not enough matching data points to compare the results for W3 and W4. Because at least the W2 offset is statistically significant, we explore whether these differences  affect our results. We rerun PROSPECTOR on the 19 galaxies in our science data set that matched galaxies having IR estimates by others. The original W1 and W2 photometry are replaced with the values from the associated matching galaxy with all other parameters unchanged.  As shown in Figure \ref{fig:WISEprospect}, the original and test results for both log(A/K) and log(sSFR$_0$)
agree well. The mean differences are $-0.16\pm0.13$ and $-0.05\pm0.12$, respectively, which are statistically insignificant offsets. We conclude that the possible offset in our W2 photometry does not affect our conclusions compared to previously published results.

\section{Determining the Minimum Necessary Photometric Data}
\label{AppSelect}

To determine the minimum photometric data needed for a reliable SED reconstruction, we use the 509 SMUDGes with acceptable photometry in all nine bands (referred to as the base group and assigned f-phot = 2 in Table \ref{tab:sfh}). By selectively removing some of the UV and IR photometry, we assess whether we can recover results that are consistent with those obtained using the full data set. The fewer bands we need, the larger our galaxy sample will be beyond the base group.

\begin{figure*}[ht]
\includegraphics[width=1.0\textwidth]{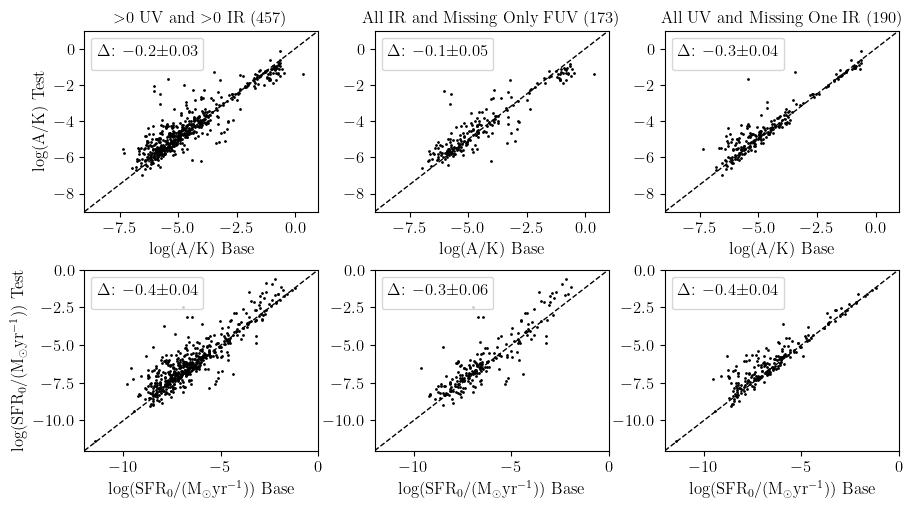}
\caption{The effect on the recovered A/K and SFR$_0$ of omitting some UV or IR observations from the 509 candidates with acceptable photometry in all nine bands (base group). In parenthesis are the numbers of additional candidates that are added to the base group for the corresponding criteria.  In each panel we present $\Delta$, the mean difference and the uncertainty in the mean between the original and the revised analysis (original $-$ revised).}
\label{fig:TestvsOrig}
\end{figure*}

We focus on cases where either the UV or IR datasets are partially incomplete. We tested three scenarios: 1) we have at least one good measurement in both UV and IR, 2) we are only missing one measurement in the UV, and 3) we are missing only one measurement in the IR.
We randomly delete bands from the base group in a way that matches the numbers of galaxies in our sample with the same missing data entry. In Figure \ref{fig:TestvsOrig} we present the comparison of the results for our two main measurements (A/K and SFR$_0$). 

It is evident from Figure \ref{fig:TestvsOrig} that the correlation remains in place (i.e. we reproduce the results obtained using all of the data, albeit with larger scatter), even when we require only one UV and one IR measurement. We opt to define this as the minimal necessary photometric data, which allows us to nearly double the sample of galaxies that we can analyze. For those who only want our most reliable results, we flag the galaxies for which all nine data bands are used in Table \ref{tab:sfh}.

\section{Setting the lower limit on the dust prior}

\begin{figure*}[ht]
\includegraphics[width=1.0\textwidth]{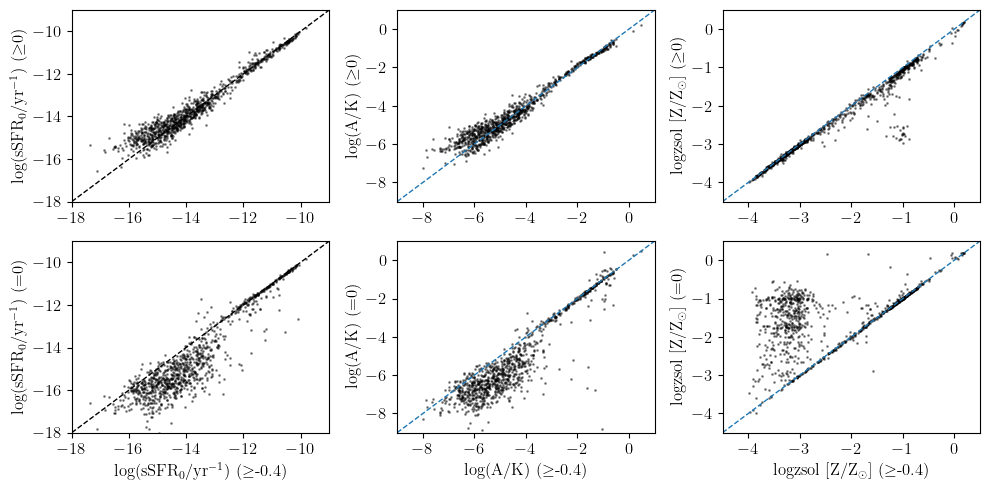}
\caption{Exploring the dependence of results on different dust priors. Upper panels compare the results when setting $A_V$ to be nonnegative vs. our choice of allowing moderately negative values. Lower panels compare the results when setting $A_V$ to 0. Effects on our key measurements of sSFR$_0$ and A/K are modest and do not affect our conclusions, but the effects on the recovered metallicity are large for quiescent galaxies.}
\label{fig:Dust}
\end{figure*}

\begin{figure*}[ht]
\includegraphics[width=1.0\textwidth]{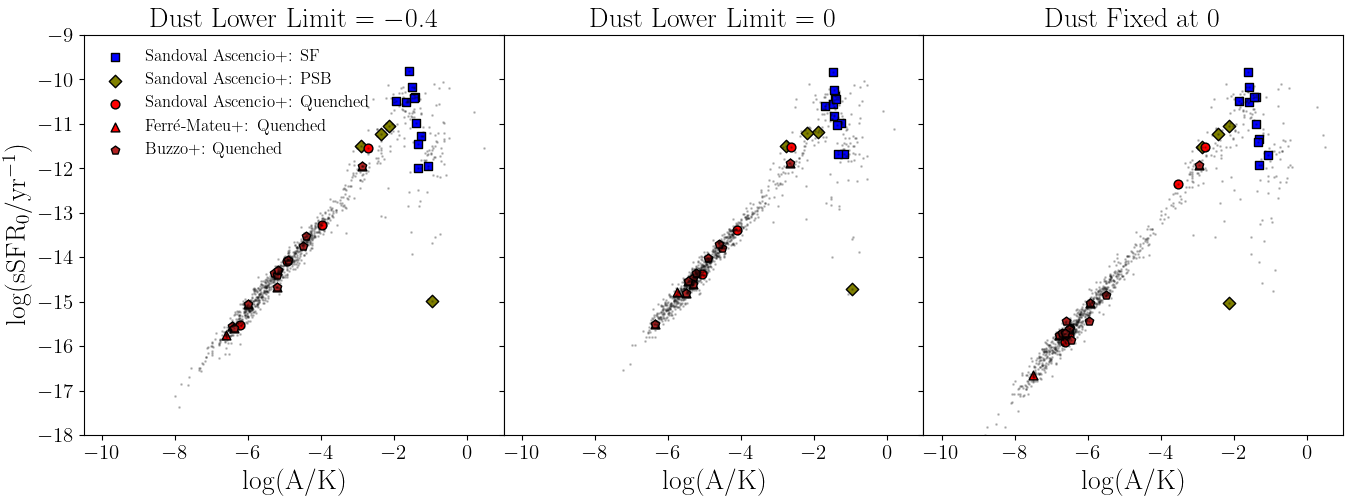}
\caption{The effects of different dust priors on one of our principle diagnostic diagrams. The leftmost panel shows results from our adopted approach, the middle panel results from setting the lower limit on A$_V$ to 0, and the rightmost panel the results from setting A$_V=0$ for all galaxies. Although there are quantitative differences, those appear to be mostly limited to the quiesent systems and our classification of the galaxies remains consistent.}
\label{fig:ClassvsDust}
\end{figure*}

As discussed in Section \ref{sec:SFH}, we intentionally set the lower limit of the dust prior to the unphysical value of $-$0.4.  Here we explore the effects of this choice on our results, particularly on our key measures of sSFR$_0$ and A/K.  As shown in Figure \ref{fig:Dust},  these parameters are nearly unchanged for larger values, particularly compared to the results of setting the lower limit of the prior to be $A_V \ge 0$ (upper panels). For smaller values, there is generally increased scatter and, in the case of setting $A_V=0$, systemic offsets. The systemic offsets in sSFR$_0$ and A/K are of little consequence since these occur for values that are already quite small. We conclude that the choice of prior, within the range explored, has no bearing on our conclusions regarding the recent star formation histories of these galaxies (see Figure \ref{fig:ClassvsDust}).

\begin{figure*}[ht]
\includegraphics[width=1.0\textwidth]{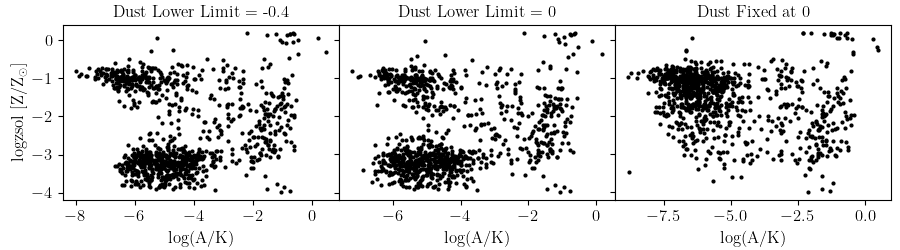}
\caption{How dust prior limits affect the recovered [Z/Z$_\odot$]. The primary difference among these results occurs for quenched galaxies (log(A/K)$<-3.5$). When $A_V$ is set to 0, the metallicity for the subgroup of these galaxies that appear to have metallicities below $-3$ move up, joining the rest of the sample at $-1$ to $-2$.}
\label{fig:logzsol}
\end{figure*}

The situation is different for metallicity determinations, which show a significant difference between when $A_V$ is allowed to vary versus when it is set to 0. Although we do not know which results are closer to reality, it seems much more likely that most of these galaxies have metallicities between $-2$ and $-1$ rather than between $-3$ and $-4$. If so, then there is a failure in the modeling, in this aspect, for some quenched galaxies when allowing for dust (Figure \ref{fig:logzsol}). We conclude that the metallicity estimates in our catalog should be treated with skepticism for galaxies with log(sSFR$_0$/yr$^{-1})< -12$ or log(A/K)$<-3$.

\label{App:Dust}

\section{PROSPECTOR SFR Continuity Condition}
\label{app:priors}
As we discussed briefly in \S\ref{sec:SFH}, we have some flexibility in assigning a prior that moderates the degree of continuity in the recovered SFRs between adjacent temporal bins.
The most cited nonparametric priors used in PROSPECTOR are referred to as the ``continuity" and ``Dirichlet" options (for specific details, see \citealt{Leja_2019}).  

As noted in Section \ref{sec:SFH}, \cite{Reyes} found that PROSPECTOR has difficulty fitting faint galaxies with the Dirichlet prior when the number of age bins is inadequate ($<6$) or the concentration factor ($\alpha_D$) is too small.  We test whether this conclusion is applicable to our dataset using simulations.  We start by selecting the 93 star-forming science galaxies with A/K $> -2$ and $-11.5 < $ sSFR$_0/$yr$^{-1} < -9.5$.  The total mass and sSFR$_0$ for each of these guide our simulations.  Our input SFH model consists of two delayed-$\tau$ components.  The first component approximates an old stellar population and starts 8 Gyr ago with an e-folding time of 3.4 Gyr and is quenched at various times in different realizations (20, 100, 200, 500, and 1000 Myr ago). We set the e-folding time to produce a log(A/K) of $-$1, approximately the location of the middle of our defined ``star-forming" region.  The second component represents recent star formation and starts at 10 Myr ago with an e-folding time of 1 Myr. For each simulation, the young component is assigned a mass equal to the mass we recovered for the youngest bin of the associated science target, and the old component's mass is set to the total mass we recovered minus this value.  We fit these simulations using the continuity  and Dirichlet priors, using four $\alpha_D$ values (0.2, 0.4, 0.7 and 1.0) for the latter. 

\begin{figure}[ht]
\begin{center}
\includegraphics[width=1.0\textwidth]{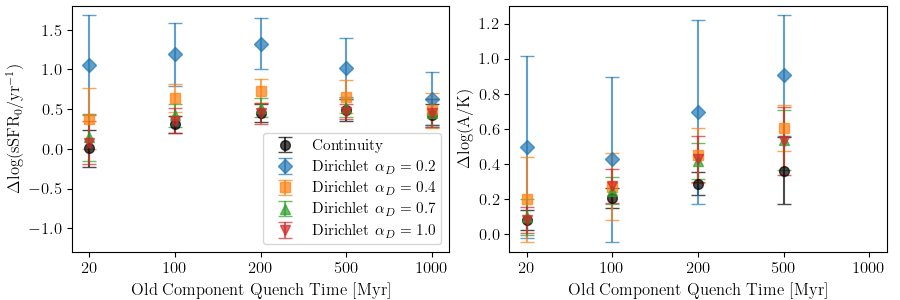}
\end{center}
\vskip -.5cm
\caption{Differences between simulation and fitted results for log(sSFR0) and log(A/K) using the continuity and four different Dirichlet priors. In all cases, there are overestimations of sSFR$_0$ and A/K. Biases for both SFR$_0$ and A/K are smaller for the continuity prior than any of the Dirichlet priors for all tested quench times. $\Delta$log(A/K) points are not shown for a quench time of 1000 Myr because the input A/K's equal zero.}
\label{fig:cont}
\end{figure}

We present our results in Figure \ref{fig:cont}, where we find that the continuity prior results in smaller errors for both SFR$_0$ and A/K than any of the Dirichlet priors for all quenching times. This supports the \cite{Reyes} conclusion that Dirichlet priors underperform when the number of age bins is small. Also, as expected from the analysis by \cite{Reyes}, Dirichlet errors increase as $\alpha_D$ decreases.  However, this finding may be limited to our particular implementation, where we use only four age bins to evaluate faint galaxies, and variations of the Dirichlet prior may be appropriate in other situations. We conclude that for our particular sample and goals, the continuity prior is preferable to the Dirichlet prior.

\section{Simulated UDGs}
\label{app:sims}

We now test the accuracy and precision with which the analysis can recover sharp changes in SFR behavior. In particular, we focus on the two measures from which we draw inferences, SFR$_0$ and A/K.  
In all cases where we refer to these quantities, even for galaxies where spectral classification exists in the literature, what we are discussing and plotting are our estimates of SFR$_0$ and A/K as we have defined these to be in terms of our PROSPECTOR model results. We  now determine the degree to which we recover our input values for simulated galaxies that are similar to our actual galaxies in specific scenarios.

\subsection{Galaxies With Young Stars}
\label{app:current_sfr}

We begin by assessing the reliability of our measure of ongoing star formation, SFR$_0$, for galaxies that are currently star-forming.
To assess the effects of the smoothing in recovered values of SFRs inherent to the continuity prior, we begin with the same two component input model used in Appendix \ref{app:priors} and the same quenching times for the old component (20 Myr, 100 Myr, 200 Myr, 500 Myr and 1 Gyr) and a young component normalized to the recovered SFR$_0$ of our actual galaxies. In Table \ref{tab:SFRrecovery} we present a summary of the results. The correlations between the input values and the recovered values are strong, indicating that the relative rankings of SFR$_0$, sSFR$_0$, and M$_*$ are preserved. However, there are systematic biases in which the output values underestimate SFR$_0$ and overestimate the stellar mass.  Our results demonstrate that we are likely underestimating SFR$_0$ by a factor of two (an average offset of $0.34\pm0.23$ dex) in any situation where there has been a pause in star formation, but that this bias is relatively insensitive to the length of that pause. This is likely due to the ``leakage" of some star formation from the youngest to the second youngest bin. This problem is not unexpected because our photometry does not present a direct constraint on stars of age $<$ 20 Myr. Adding, H$\alpha$ photometry, for example, could address this shortcoming. If star formation has not recently paused, then we recover the input value, presumably because the leakage between the adjacent bins comes close to averaging out. These differences are negligible on the scale on which we classify our galaxies (e.g. Figure \ref{fig:categorize}). We conclude that we recover sSFR$_0$ well, to a factor of two or better.

We perform a similar analysis on A/K using the same simulations.  In this case, we compare simulations in which we set the e-folding time of the old component to produce a log(A/K) of $-1$ if the quenching time is more recent than 150 Myr ago.  We obtain log(A/K) differences of $0.08\pm0.06$, $0.21\pm0.06$, $0.29\pm0.07$, and $0.36\pm0.19$ dex, respectively, for quenching times of 20, 100, 200, and 500 Myr.  We do not include the simulations with a quenching time of 1 Gyr since these contain no mass in the next to oldest age bin (i.e. log(A/K) $= -\infty$).  As expected, when quenching occurs at 500 Myr (within the bin), log(A/K) is reduced. The mean value is now $-$1.21.      

Finally, in Figure \ref{fig:ssfr} the lack of star forming systems with low values of A/K is striking and naively suggests that old stellar populations do not suddenly restart star formation. 
Identifying systems in which a purely old stellar population suddenly restarts star formation is difficult for various reasons. First, the recent star formation would have to confine itself to the latest 100 Myr or otherwise it would extend into the bin we use to define intermediate age stars. This would require us to be rather fortuitous in finding such systems. Second, even such a system (one with star formation only in the recent past) will have formed intermediate mass stars that the analysis might have difficulty distinguishing from an intermediate age population (for example, if the initial mass function is not quite correct). Finally, --- and the issue we are discussing here --- our analysis will tend to smooth out the SFH and, therefore, avoid placing galaxies in this region of the diagram.
Here we explore whether our analysis could have missed such a population of young stars in otherwise quiescent galaxies. 

\begin{deluxetable}{lrrr}
\tablewidth{\columnwidth}
\tablecaption{Comparisons between simulated input and output values for star-forming galaxies$^a$ }
\label{tab:SFRrecovery}
\tablehead{
\colhead{Quench time}&
\colhead{$\Delta\log({\rm sSFR}_0/{\rm yr}^{-1})$} &
\colhead{$\Delta\log({\rm SFR}_0/({\rm M}_\odot {\rm yr}^{-1}))$} &
\colhead{$\Delta\log({\rm M}_*/{\rm M}_\odot$)}\\
\colhead{[Myr]}& & & 
}
\startdata
20   & $0.01\pm0.24$ (r = 0.85) & $-0.11\pm0.24$ (r = 0.97) & $-0.12\pm0.01$ (r = 1.00)\\
100  & $0.31\pm0.11$ (r = 0.97) & $ 0.18\pm0.11$ (r = 0.99) & $-0.13\pm0.01$ (r = 1.00)\\
200  & $0.45\pm0.11$ (r = 0.99) & $ 0.32\pm0.11$ (r = 0.99) & $-0.13\pm0.01$ (r = 1.00)\\
500  & $0.49\pm0.14$ (r = 0.99) & $ 0.36\pm0.12$ (r = 0.99) & $-0.13\pm0.02$ (r = 1.00)\\
1000 & $0.43\pm0.13$ (r = 1.00) & $ 0.33\pm0.12$ (r = 1.00) & $-0.10\pm0.02$ (r = 1.00)\\
 \enddata
\centering{
\tablenote{Values are for output minus input. The r values given in parentheses are the Pearson correlation coefficient and demonstrate that the relative rankings remain nearly unchanged.}}
\end{deluxetable}
\onecolumngrid

\begin{figure}[ht]
\begin{center}
\includegraphics[width=0.5\textwidth]{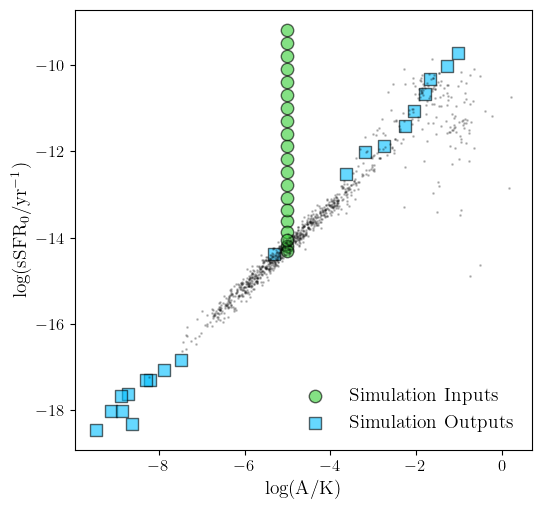}
\end{center}
\vskip -.5cm
\caption{Simulating 16 galaxies with long standing quiescent populations and different levels of current star formation. We conclude that such systems are not correctly recovered in our treatment.}
\label{fig:simLoAK}
\end{figure}

As above, we address this question with simulations using two component delayed tau models.
We again fix dust extinction at 0, $logzsol$ at $-1$, total mass at 10$^8$M$_\odot$ and redshift at 0.01.  We set the old component to produce a log(A/K) of $-$5 with no quenching which simulates the value in about the middle of the quiescent region of our science targets.  The recent star-forming component starts at 10 Myr with an e-folding time of 1 Myr with the only variable being the mass of this component.  We run 16 simulations starting with the mass of the science target with the largest sSFR$_0$ in the sample and reducing this by 50\% for each subsequent simulation. As shown in Figure \ref{fig:simLoAK}, the PROSPECTOR results differ significantly from the input parameters and do not reproduce the condition where there is recent sudden star formation after a long period of quiescence.  We conclude that we cannot say with confidence that the lack of SMUDGes in this region of the diagram reflects an actual lack of such galaxies.

\subsection{Galaxies with Intermediate Age Stars}

In Figure \ref{fig:ssfr} there is a reversal in the relationship between sSFR$_0$ and A/K for log(A/K) $> \sim -2$.  This behavior is qualitatively what one might expect for post-starburst systems. In such galaxies, the current SFR declines, and the fraction of A stars increases.
However, as always, we need to determine if the modeling is indeed faithfully recovering the behavior. 

We explore whether galaxies found in the downturn seen in Figure \ref{fig:ssfr} can be reproduced with a simple scenario or are necessarily ``scattered" into this region of the diagram. As previously done, we create models consisting of two delayed-$\tau$ components. The first  simulates an old population and starts at 8 Gyr with an e-folding time of 3.4 Gyr, which produces log(A/K) of $-$1 and approximates the location of the middle of the downturn of the science targets (Figure \ref{fig:ssfr}). Unlike in Appendix \ref{app:current_sfr}, we now simply use a fixed stellar mass of 10$^8$ $M_\odot$. We simulate the post-starburst effect by truncating star formation in this component at either 100 or 200 Myr (corresponding to the edges of our second and third age bins).  The second component simulates a burst starting at 10 Myr with an e-folding time of 0.5 Gyr and a variable mass that approximates the sSFRs of our science targets. Specifically, we use the 16 targets with $-1.2 <$ log(A/K) $< -0.8$ and sSFR$_0$ $< -11.5$.  Because these are already selected to be in the downturn region they have lower SFR$_0$ than the star forming galaxies we examined in Appendix \ref{app:current_sfr}. All other parameters are kept constant with dust extinction set to 0, $logzsol$ to $-1$, and redshift to 0.01. We set the uncertainties in flux using those of our science targets with similar flux values and the resulting photometry is processed identically as for our science targets.  

\begin{figure*}[ht]
\begin{center}
\includegraphics[width=1.0\textwidth]{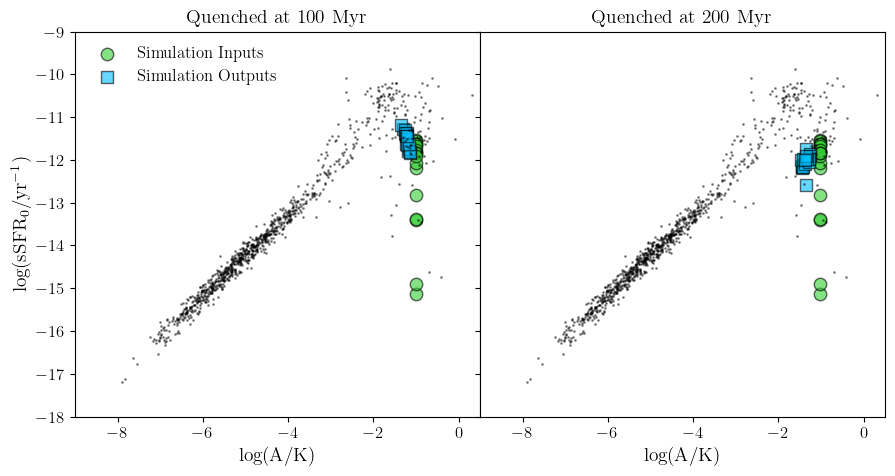}
\end{center}
\caption{Testing the downturn in the log(sSFR$_0$) vs. log(A/K) relationship. Sixteen simulated galaxies, constructed to match the observed properties of galaxies in the downturn, are modeled and passed through our analysis. The output values tend to slightly lower values of log(A/K) and log(sSFR$_0$) but are generally still like in the downturn region of Figure \ref{fig:ssfr}, confirming that galaxies with recently quenched star formation are identified as such even if they have a modest level of residual star formation.
}
\label{fig:sims}
\end{figure*}

In Figure \ref{fig:sims} we present the results of these simulations, where we find that we qualitatively replicate the downturn. Nevertheless, we do find some offsets.  The resulting sSFR$_0$'s are generally larger than those input, particularly when the input sSFR$_0$ is exceedingly low, probably due to ``leakage" into the youngest age bin. Moreover, when  the quenching is set to 200 Myr, which allots no mass to the second youngest age bin, there is a more significant decrease in log(A/K), presumably caused by leakage from the second oldest bin into that bin.  We are encouraged by these findings with respect to our ability to distinguish recent declines in SFRs.

\end{appendix}

\end{document}